\documentstyle[12pt,psfig,axodraw]{article}

\catcode`@=11
\long\def\@caption#1[#2]#3{\par\addcontentsline{\csname
  ext@#1\endcsname}{#1}{\protect\numberline{\csname
  the#1\endcsname}{\ignorespaces #2}}\begingroup
    \small
    \@parboxrestore
    \@makecaption{\csname fnum@#1\endcsname}{\ignorespaces #3}\par
  \endgroup}
\catcode`@=12

\input{epsf}
\ifx\epsffile\undefined
\def\epsffile#1#2#3#4]#5{}
\fi
\input{rotate}

\newcommand{\beq}{\begin{equation}}   
\newcommand{\eeq}{\end{equation}}
\newcommand{\bear}{\begin{eqnarray}}
\newcommand{\eear}{\end{eqnarray}}
\newcommand{\ba}{\begin{array}}      
\newcommand{\ea}{\end{array}}

\newcommand{\gsim}{\lower.7ex\hbox{$\;\stackrel{\textstyle>}{\sim}\;$}}
\newcommand{\lsim}{\lower.7ex\hbox{$\;\stackrel{\textstyle<}{\sim}\;$}}

\newcommand{\met}{{\not\!\! E_T}}
\newcommand{\mchi}{{m_{\bf n}}}
\def\VEV#1{\left\langle #1\right\rangle}
\def\ap  #1 #2 #3 #4 {Ann.~Phys.         {\bf  #1}, #2 (#3)#4 }
\def\aplb#1 #2 #3 #4 {Acta Phys.~Pol.    {\bf B#1}, #2 (#3)#4 }
\def\cpc #1 #2 #3 #4 {Comp.~Phys.~Comm.  {\bf  #1}, #2 (#3)#4 }
\def\epjc#1 #2 #3 #4 {Eur.~Phys.~J.      {\bf C#1}, #2 (#3)#4 }
\def\jetp#1 #2 #3 #4 {JETP Lett.         {\bf  #1}, #2 (#3)#4 }
\def\npb #1 #2 #3 #4 {Nucl.~Phys.        {\bf B#1}, #2 (#3)#4 }
\def\plb #1 #2 #3 #4 {Phys.~Lett.        {\bf B#1}, #2 (#3)#4 }
\def\prd #1 #2 #3 #4 {Phys.~Rev.         {\bf D#1}, #2 (#3)#4 }
\def\prl #1 #2 #3 #4 {Phys.~Rev.~Lett.   {\bf  #1}, #2 (#3)#4 }
\def\ptp #1 #2 #3 #4 {Prog.~Theor.~Phys. {\bf  #1}, #2 (#3)#4 }
\def\zpc #1 #2 #3 #4 {Zeit.~Phys.        {\bf C#1}, #2 (#3)#4 }

\begin{document}

\pagestyle{empty}
\begin{titlepage}
\def\thepage {}        

\title{\bf New Higgs Signals from Flavor Physics \\ [2mm]
           in Large Extra Dimensions \\ [1cm]}

\author{ {\small \bf Hsin-Chia Cheng
                 and Konstantin T.~Matchev} \\ 
\\
{\small {\it Theoretical Physics Department}}\\  
{\small {\it Fermi National Accelerator Laboratory}}\\
{\small {\it Batavia, Illinois, 60510, USA \thanks{e-mail
  addresses: hcheng@fnal.gov, matchev@fnal.gov} }}\\ }

\date{ }

\maketitle

   \vspace*{-9.4cm}
\noindent
\makebox[10.7cm][l]{August 11, 1999} FERMILAB-PUB-99/216-T \\ [1mm]

\vspace*{8.5cm}

\baselineskip=18pt

\begin{abstract}

{\normalsize In a realistic theory with large extra dimensions and TeV
scale gravity, there are often other particles living in the extra
dimensions in addition to gravitons. A well motivated candidate
is the flavon in a flavor theory which explains the smallness
of the fermion masses by the large volume of the extra dimensions.
The flavon interactions involve the Higgs field and can therefore
give rise to new mechanisms of Higgs boson production.
We study these new Higgs signals, including both real flavon
emission and virtual flavon exchange. They could provide
powerful probes of the string scale which are comparable
to the extensively studied graviton production, if
the flavons and the graviton feel the same number of
extra dimensions. If the flavons only live in a subspace of
the extra dimensions, they can yield even stronger bounds and
may generate spectacular displaced vertex signals.
Our study also applies to a four dimensional flavor theory
with light flavons as well as to any theory with bulk scalar
fields coupled to the Higgs and fermions.}

\end{abstract}

\vspace{0.5cm}
\centerline{\sl Submitted to Nucl. Phys. B}

\vfill
\end{titlepage}

\baselineskip=18pt
\pagestyle{plain}
\setcounter{page}{1}

\section{Introduction}
\label{introduction}
\setcounter{equation}{0}

The Standard Model (SM) is very successful in explaining
most precision data from current collider experiments,
but still we have not found (or tested) all ingredients of the SM. 
The elusive Higgs particle is yet to be discovered, and
its interactions with the SM fermions and gauge bosons
will need to be measured before we close the final chapter in the
quest for the correct theory of particle physics at the weak scale.

Given the success of the Standard Model, there are still several
unresolved puzzles. An important one is related to 
flavor, i.e. the size of the fermion masses
and the generation of the CKM matrix. In the Standard Model,
the fermion masses arise from the Yukawa interactions between
the fermions and the Higgs. We do not understand why all
fermion Yukawa couplings other than the top Yukawa coupling are 
so small or why the CKM matrix is approximately diagonal.
The smallness of the fermion Yukawa couplings also leads to
the suppression of direct $s$-channel Higgs production, 
making the Higgs discovery a challenging task.

A natural explanation for the smallness of the fermion Yukawa
couplings is that there is an enlarged (flavor) symmetry under
which the direct couplings between the light fermions and the Higgs
are forbidden in the symmetry limit. Only after the flavor symmetry
is broken by the vacuum expectation value (vev) of some scalar
flavon field $\langle \varphi \rangle$, the Yukawa couplings of
light fermions are generated by some higher dimensional operators
which are invariant under the flavor symmetry, such as
\beq
\label{hdop}
{y_f \over M_F} \varphi H \bar{f}_L f_R,
\eeq
with $y_f \sim 1$.
The Yukawa coupling is given by $y_f \VEV{\varphi}/M_F$ and is small
if $\VEV{\varphi} \ll M_F$. For a more general flavor theory, there
can be several $\varphi$'s and the Yukawa couplings can be
suppressed by higher powers of $\VEV{\varphi}/M_F$. The higher
dimensional operators like (\ref{hdop}) can arise from 
gravitational scale physics, in which case $M_F$ will be the 
Planck scale, or from integrating out some vector-like
Froggatt-Nielsen fields~\cite{FN}, in which case $M_F$ will
be the mass of the Froggatt-Nielsen fields. The scale of the
flavor physics is not determined by (\ref{hdop}) as long as
the ratio $\VEV{\varphi}/M_F$ produces the necessary fermion
mass hierarchy. However, the flavor symmetry breaking scale
in the usual four dimensional theory is typically quite high. 
If the flavor symmetry is a continuous global
symmetry, there will be one or more massless Goldstone bosons,
in which case there are strong constraints from experiments and
astrophysics requiring the symmetry breaking scale to be greater than
$\sim 10^9$ GeV. In the cases of a gauge symmetry or a discrete
symmetry, the gauge bosons and the flavons will induce flavor-changing
effects, depending on the symmetry and its breaking pattern.
On the other hand,
the absence of small couplings in the fundamental theory implies that
the masses of the flavons are on the order of (or smaller than)
their vevs. 
Since the masses are constrained from experiment
(the non-observation of new gauge bosons and 
compliance with rare flavor-changing processes \cite{CH,ACHM})
to be well above the weak scale, the flavon vevs themselves must
lie in that range too. Then, to get a small ratio of
$\VEV{\varphi}/M_F$, one has to consider values of $M_F$
much larger than the weak scale.

Another unsettling issue with the SM is the hierarchy
problem. A few proposals (supersymmetry, technicolor)
have been around for a long time, but the most radical approach
\cite{ADD} was put forth only recently. The idea is to
introduce extra large compact spatial dimensions to
bring the fundamental physics scale down to a few TeV,
thus ameliorating the hierarchy problem. 
Without the large hierarchy between the fundamental Planck scale
and the weak scale, any flavor physics must occur close
to the weak scale. In fact, without the grand ``desert'' to
protect the SM from various problems, such as proton decay
and large flavor-changing effects, it is desirable to have
a large (discrete) flavor symmetry to prevent these disastrous 
large effects~\cite{AD}. The hierarchy of the fermion masses
in this framework can originate from the large size of the 
extra dimensions instead.
For example, the flavor symmetry may be maximally broken on one or
several distant branes in the extra dimensions. The breaking is then
transmitted to the brane where the SM fields reside by some
bulk fields with the same quantum numbers as the flavor
symmetry breaking fields on the distant branes.
Depending on the distances and the masses of the bulk fields,
the symmetry breaking effects on our brane may be suppressed
by powers of or exponentially with the distance, which explains
the smallness of the Yukawa couplings~\cite{AD}.
Specifically, the interactions on our brane look like
\beq
\label{chi}
{y'_f \over M_s^{1+\delta/2}} \int d^4 x {\bar f}_L f_R H \chi(x,y=0),
\eeq
where $\chi$ is the bulk messenger field (with mass dimension $1+\delta/2$,
$\delta$ is the number of the extra dimensions accessible to $\chi$),
$y$ represents the coordinates of the extra dimensions, and $M_s$ is the
fundamental scale of the higher-dimensional theory, possibly
the string scale. $\chi$ has a nonzero ($y$ dependent) vev 
induced from the distant brane. After propagating the
distance between the two branes, on our brane $\chi$ has a
vev $\VEV{\chi(y=0)}\ll M_s^{1+\delta/2}$ and as a result, the corresponding
Yukawa coupling $\lambda_f = y'_f \VEV{\chi(y=0)}/ M_s^{1+\delta/2}
\ll 1$.
In this framework, the smallness of $\VEV{\chi(x,y=0)}$ comes from
the distance suppression and does not require the mass of
$\chi$ to be of order its vev $\VEV{\chi}$, in
contrast to a four dimensional theory.

The string scale is expected to be close to the Planck scale in
the higher dimensional theory. For $n$ extra dimensions accessible
to the gravitons ($n\geq \delta$), the $4+n$ dimensional Planck
scale $M_*$ and reduced Planck scale $\hat{M}_*$ are related to
the 4 dimensional Newton's constant $G_N$ by~\cite{ADD}
\beq
\hat{M}_*^{2+n} V_n = M_*^{2+n} V_n/(2\pi)^n = 
(4\pi G_N)^{-1} \equiv \hat{M}_P^2,
\eeq
where $V_n$ is the $n$ dimensional volume of the extra dimensions
($V_n=L^n$ if all extra dimensions have the same length $L$).
To avoid retaining too many similar scales and to simplify
the notation, we will use the $4+n$ dimensional reduced Planck
scale for most of the paper and define
\beq
M \equiv \hat{M}_*.
\eeq
The interaction (\ref{chi}) can be rewritten
in terms of $M$,
\beq
\label{chiint}
{y_f \over M^{1+\delta/2}} \int d^4 x {\bar f}_L f_R H \chi(x,y=0).
\eeq
Remember that gravitons have order one couplings in units of
the reduced Planck scale. We will assume $y_f\sim 1$ for our study.

The bulk messenger field $\chi$ (which from now on we shall call a ``flavon'',
in analogy to the case of a four dimensional flavor theory)
can live in a subspace of $4+\delta$ dimensions with $\delta \leq n$.
{}From a 4 dimensional point of view, the $\chi$ field
is Fourier decomposed into a Kaluza-Klein (KK) tower of states,
\beq
\chi(y=0)={1\over \sqrt{V_\delta}} \sum_{\bf n} \chi_{\bf n},
\eeq
where $V_\delta$ is the volume of $\delta$ extra dimensions
($V_\delta=L^{\delta}$ for equal length), and
${\bf n}$'s are integer vectors in the $\delta$ dimensions.
The operator (\ref{chiint}) can then be written as
\beq
\label{KKint}
{y_f \over M_F} \sum_{\bf n} {\bar f}_L f_R H \chi_{\bf n},
\eeq
where we define
\beq
M_F \equiv\ M^{1+\delta/2} V_\delta^{1/2},
\label{Mdef}
\eeq
so that the interaction has the same form as (\ref{hdop}).
If $\chi$ propagates in the same extra dimensions as gravity,
i.e. if $\delta=n$,
$M_F$ can be identified with the 4-dimensional reduced Planck scale
$\hat{M}_P=(4\pi G_N)^{-1/2}$.
However, $M_F$ can in principle lie anywhere between
$M$ and $\hat{M}_P$. For example, if all extra dimensions
have the same length $L$, we find
$M_F=M(\hat{M}_P/M)^{\delta/n}=\hat{M}_P(M/\hat{M}_P)^{1-\delta/n}$.

Although each interaction (\ref{KKint}) is highly
suppressed by the large scale $M_F$, the
large number of KK states compensates for the 
suppression. In the end of the day,
the cross-sections will only depend of $M$ and $\delta$ and
not on $M_F$ or $L$, which we only introduce for notational clarity.

For very light bulk messengers, i.e. ``flavons'',
(with zero mode mass $m_0<1$ GeV), there are various
constraints from astrophysics and flavor violating effects
pushing the lower bound of $M$ to be far beyond the weak
scale for $\delta \leq 4$ \cite{AD}. All these constraints 
disappear if $m_0$ is greater than a few GeV, because the
$\chi$ KK modes would be too
heavy to be produced in these processes. In that case,
$M$ is only constrained by the graviton KK state interactions
and can be $\sim$ TeV for $n > 2$. For such a low $M$,
the interaction (\ref{chiint}) can be important at high energy
colliders and give rise to new Higgs signals in future experiments.

The possibility of $\chi$ living in fewer dimensions than gravity 
may result in some advantages of collider probes of the $\chi$ fields
compared to the more model independent signals from KK gravitons.
First, the resulting limits on the gravitational scale in collider
experiments are stronger for smaller number of extra
dimensions~\cite{GRW,MPP}. However, the existing lower
bounds on $M$ are also stronger for smaller $n$ ($n=1$ is ruled
out, $M_*>50$ TeV from SN1987A~\cite{SN1987A} and $M_*>110$ TeV
from cosmic diffuse gamma ray radiation for $n=2$~\cite{HS}). 
As mentioned above, these
constraints do not apply to $\chi$ if $\chi$ is heavier than a few
GeV. It is possible that $n$ is large, yet $\delta$ is small,
so that $M$ is quite low and we have larger collider signals for
$\chi$ rather than the KK states of the graviton.
Second, because $M_F$ can be smaller than $\hat{M}_P$ for $\delta<n$,
there is a range of $M_F$ where the KK modes of $\chi$ fields
can have easily detectable macroscopic decay lengths inside the detector.

The interactions between different fermions and $\chi$'s depend
on the flavor symmetry and how it is broken. A large flavor symmetry
is favored to avoid the flavor violation problem coming from higher
dimensional operators generated at the string scale $M_s$.
A straightforward example
for the flavor symmetry is (a large discrete subgroup of) 
U(3)$^5$~\cite{AD,Hall}. The $\chi$ fields transform as $(3,\,
\bar{3})$ under the product of two U(3)'s. The vevs of $\chi$'s
break U(3)$^5$ and generate the SM fermion Yukawa couplings. It is
known that the flavor changing effects are safe if the SM fermion
Yukawa couplings are the only source of U(3)$^5$ breaking~\cite{CG}.
Additional flavor violating effects may be induced from the $\chi$
interaction away from our brane but are suppressed by distances~\cite{Hall}.
It provides a working example and for the most part of this
paper we shall work under this assumption, that is, we will
assume that there is one and only one $\chi$ field for each
entry of the SM fermion Yukawa matrices, e.g., $H\bar{q}_iq_j\chi_{q_iq_j}$
or $H\bar{\ell}_i\ell_j\chi_{\ell_i\ell_j}$. For a more general flavor 
theory, there may be $\chi$ fields which couple to more than one
combinations of fermions,
and the Yukawa couplings may come from interactions
containing higher powers of $\chi$ fields or products
of different $\chi$'s.

To summarize, the scenario that we are considering here
is completely determined by just a few independent
parameters (assuming that all extra dimensions have the same size):
$$ 
\left\{ M,m_0,m_h,\delta, n\right\}.
$$ 
$M_F$ and $L$ are derived quantities:
$M_F=M(\hat{M}_P/M)^{\delta/n}$ and
$L=M^{-1}(M_P/M)^{2/n}$. The allowed range of $M$ and $n$
is constrained by the astrophysical and collider bounds
on gravity in large extra dimensions. In Table~\ref{limits}
we list the current limits on the scale $M$ from real graviton
emission at LEP and the Tevatron, translated from the studies
of several groups~\cite{GRW,MPP,CK,BHRYD}. We do not include bounds
from processes with virtual gravition exchange, where
\begin{table}[t!]
\centering
\renewcommand{\arraystretch}{1.5}
\begin{tabular}{||c|c||c|c|c|c|c||}
\hline\hline
Ref. & Process & \multicolumn{5}{c||}{Number of extra dimensions}     \\ 
\cline{3-7}
     &         &    2     &    3     &     4    &     5    &    6     \\ 
\hline\hline
\cite{GRW}  & $e^+e^-\rightarrow \gamma G$
               &   522    &   324    &  231     &    178   &   ---    \\ \hline
\cite{MPP}  & $e^+e^-\rightarrow \gamma G$
               &   478    &   ---    &  214     &    ---   &   131    \\ \hline
\cite{CK}   & $e^+e^-\rightarrow \gamma G$
               &   318    &   ---    &  ---     &    ---   &   ---    \\ \hline
\cite{BHRYD}& $e^+e^-\rightarrow Z G$
               &   372    &   156    &   91     &    ---   &   ---    \\ \hline\hline
\cite{GRW}  & $p\bar{p}\rightarrow j G$
               &   522    &   362    &  280     &    208   &   ---    \\ \hline
\cite{MPP}  & $p\bar{p}\rightarrow j G$
               &   300    &   ---    &  179     &    ---   &   154    \\ \hline
\cite{BHRYD}& $p\bar{p}\rightarrow W G$
               &   355    &   ---    &  228     &    ---   &   169    \\ \hline
\cite{BHRYD}& $p\bar{p}\rightarrow Z G$
               &   304    &   ---    &  211     &    ---   &   179    \\ \hline\hline
\end{tabular}
\parbox{5.5in}{
\caption{Current limits on the scale $M$ (in GeV) 
from real graviton emission processes,
for different number $n$ of extra dimensions.
\label{limits}}}
\end{table}
the appearance of an explicit ultraviolet cutoff in the calculation
makes it difficult to interpret the limit on $M$.
As we mentioned earlier, $m_0$ has to be larger
than a few GeV, while $m_h$ has to satisfy the experimental
limit from the Higgs search.

The plan of the paper is as follows. In Section~\ref{conventional}
we first check whether the presence of the additional operator
(\ref{KKint}) can affect the conventional Higgs
searches at LEP and the Tevatron. In the next two sections we
explore the new discovery signatures offered by the coupling
(\ref{KKint}). In Section~\ref{direct} we concentrate on 
the direct production of KK states
$\chi_{\bf n}$, leading to rather unorthodox Higgs signals
associated with missing energy or displaced vertices.
Then in Section~\ref{virtual} we
consider new Higgs production processes with a virtual exchange
of KK states $\chi_{\bf n}$.
In all cases, we estimate the discovery potential of both LEP and the
Tevatron in the new channels, and also comment on the 
possible reach of LHC or NLC. We summarize and draw our
conclusions in Section~\ref{conclusions}.

\section{Conventional Higgs searches}
\label{conventional}

The traditional searches
\cite{Run II Higgs studies, LEP Higgs searches}
look for the SM Higgs in associated
production $f\bar{f}\rightarrow Wh, Zh$ and gluon fusion
$gg\rightarrow h$. A two-Higgs doublet model, e.g., the MSSM,
also presents the opportunity to look for
$p\bar{p}\rightarrow hb\bar{b}$ at large values of $\tan\beta$.
The specific final state topology depends on the Higgs mass $m_h$.
For $m_h \lsim 140$ GeV, the Higgs decays predominantly to
$b\bar{b}$ pairs, while for $m_h \gsim 140$ GeV, the decays to
$W^+W^-$ pairs dominate. All of these searches will be
significantly affected if the higher dimensional couplings to
flavons can change dramatically the branching ratios
of the Higgs to $b\bar{b}$ or $W^+W^-$. Indeed, $\Gamma(h\rightarrow b\bar{b})$
is quite small, which has already prompted suggestions to look for other
(e.g. invisible \cite{MW,InvH}) Higgs decay modes for low Higgs masses.
At the same time, the Higgs search is the cornerstone of the Tevatron Run II
physics program, peaking in sensitivity for a range of Higgs 
masses just beyond the LEP-II reach.
As for somewhat heavier Higgs masses, $m_h\gsim 140$ GeV,
the width $\Gamma(h\rightarrow W^+W^-)$ is already rather large,
so the branching ratios to gauge boson pairs are unlikely to be
affected by any new physics.

\begin{figure}[ht]
\centerline{\psfig{figure=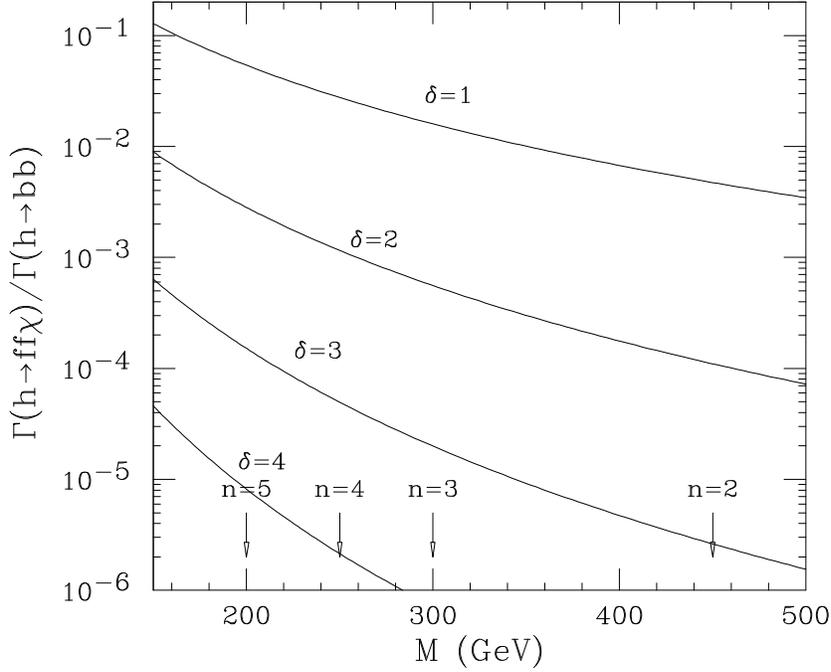,height=3.5in}}
\begin{center}
\parbox{5.5in}{
\caption[] {\small The ratio of the Higgs widths
$\Gamma(h\rightarrow \chi f\bar{f})/\Gamma(h\rightarrow b\bar{b})$,
as a function of $M$, for different values of $\delta$, 
and for fixed $m_h=120$ GeV, $m_0=10$ GeV, $N_c=3$
and $y_f=1$. The arrows are to remind the reader
of the typical limits on $M$ from graviton emission processes.
\label{width_ratio}}}
\end{center}
\end{figure}

In order to estimate the effect of the operator
(\ref{KKint}) on the Higgs branching ratios, we
have to compare the width $\Gamma(h\rightarrow b\bar{b})$
to $\Gamma(h\rightarrow \chi jj)$ or 
$\Gamma(h\rightarrow \chi \ell^+\ell^-)$.
To leading order in the fermion mass $m_f$, one has
\beq 
\Gamma(h\rightarrow f\bar{f})\ =\ 
{N_c\, g^2 m_f^2\over 32\pi M_W^2}\ m_h,
\eeq
where $N_c$ is the number of colors of $f$ and $M_W$ is the $W$ mass.
In the presence of (\ref{chiint}), and for $m_0<m_h$,
three-body decays to a fermion
pair and the corresponding flavon are also allowed. For a single
KK mode of $\chi$,
\beq
\Gamma(h\rightarrow \chi_{\bf n} f\bar{f})\ =\ 
{N_c\, y_f^2 m_h^2\over 256\pi^3 M_F^2}\ m_h\,
\biggl\{{1\over6}(1+9\mu-9\mu^2-\mu^3)+\mu(1+\mu)\log\mu\biggr\},
\eeq
where $\mu\equiv m_{\bf n}^2/m_h^2$ and $m_{\bf n}$
is the $\chi_{\bf n}$ mass.
Summing over the KK modes we have
\bear
\Gamma(h\rightarrow \chi f\bar{f}) &=&
\int_{m_0^2}^{m_h^2} d m_{\bf n}^2 \rho(m_{\bf n}) 
\Gamma(h\rightarrow \chi_{\bf n} f\bar{f}) \\ \nonumber
&=& {1\over (4\pi)^{\delta/2} \Gamma(\delta/2)}\,
 {N_c y_f^2\over 256\pi^3}\, {m_h^{\delta+3}\over M^{\delta+2}}\,
 G_{\delta}(\mu_0),
\eear
where $\rho(m_{\bf n})$ is the density function of the KK modes
given in the Appendix, and we have defined $\mu_0=m_0^2/m_h^2$ and
\beq
G_{\delta}(\mu_0)=\int_{\mu_0}^1 d \mu (\mu-\mu_0)^{(\delta-2)/2}
\biggl\{{1\over6}(1+9\mu-9\mu^2-\mu^3)+\mu(1+\mu)\log\mu\biggr\}.
\eeq
In the limit of $m_0\rightarrow 0$, the $G_{\delta}(0)$ integral can
be performed analytically and the result is
\beq
G_{\delta}(0)= {128\over \delta (2+\delta)^2 (4+\delta)^2 (6+\delta)},
\eeq
which ranges from $8.1\times 10^{-2}$ for $\delta=1$ to
$2.8\times 10^{-4}$ for $\delta=6$.
The ratio of the widths of the new $h\rightarrow \chi f\bar{f}$
channel to $h\rightarrow b\bar{b}$ is then given by
\bear
{\Gamma(h\rightarrow \chi f\bar{f}) \over
 \Gamma(h\rightarrow b\bar{b})}\ &=&\  {N_c\over3}\
{1\over 8\pi^2}\ {y_f^2\over g^2}\ 
\Biggl({ M_W\over m_b }\Biggr)^2\ 
\Biggl({m_h \over M}\Biggr)^{\delta+2}
{G_{\delta}(\mu_0) \over (4\pi)^{\delta/2}\Gamma(\delta/2)}\\ \nonumber
&\sim &\
3.0\, N_c\, {G_{\delta}(\mu_0) \over (4\pi)^{\delta/2}\Gamma(\delta/2)}\,
{y_f^2\over g^2}\ 
\Biggl({m_h\over M}\Biggr)^{\delta+2}
\label{width ratio}
\eear
We show this ratio in Fig.~\ref{width_ratio} for fixed $m_h=120$ GeV,
$m_0=10$ GeV, $N_c=3$ and $y_f=1$ (for larger $m_0$, it will be even smaller). 
In the figure we also mark the typical limits from Table~\ref{limits}
on the scale $M$ for various values of $n$.
We see that the branching ratio to flavons is typically
orders of magnitude smaller than the branching to $b\bar{b}$,
and is therefore unlikely to have any impact on the
traditional Higgs searches\footnote{Our conclusion in this respect
differs from that of Ref.~\cite{AD}, where the numerical factor
$G_\delta(\mu_0)/\left[8\pi^2(4\pi)^{\delta/2}\Gamma(\delta/2)\right]$
was not included.}. The only exception is the case of
$\delta=1$, where it can be as large as a few percent.
One should not forget that the width being compared to
$\Gamma(h\rightarrow b\bar{b})$ in Fig.~\ref{width_ratio}
is for a single flavon decay mode. In principle, one should sum over
all flavon flavors, which give similar final state signatures, e.g.
Higgs decays to two jets plus missing energy can be obtained
from decays to {\em any} $\chi_{q_i q_j}, \ q_i,q_j=u,d,s,c$. 
This will enhance the branching fraction, but one loses
the b-tagging option, which is crucial for reducing the
background in a hadronic environment. On the other hand, there
is also the possibility for decays to leptons\footnote{From
this point on we use the term ``lepton'' ($\ell$)
to denote a muon or electron only.} 
via $h\rightarrow \ell\ell'\chi_{\ell\ell'}$, albeit
with a much smaller branching fraction
(by a factor of 7.3) than $h\rightarrow jj\chi_{qq}$.
The final state is indistinguishable from the SM decay
$h\rightarrow \tau\tau\rightarrow 
\ell \nu_\ell\ell'\bar{\nu}_{\ell'}\nu_\tau\bar{\nu}_\tau$,
which already has a branching ratio of about 0.6\%,
much larger than what we typically would expect
for $h\rightarrow \ell\ell'\chi_{\ell\ell'}$.
Since the tau decays are uncorrelated, the different flavor
decays ($\ell\ne\ell'$) are not particularly distinctive either,
besides, the major backgrounds from $t\bar{t}$ and $W^+W^-$
half of the time also yield different flavor leptons.

In conclusion of this section,
we see no easy way to discover or confirm this
scenario by looking at the decay modes of the Higgs particle. 
One (rather exotic) option for that would be 
to build a muon collider at the Higgs mass and then measure
precisely the tau decay modes of the Higgs, looking for
nonuniversalities. Even then, on has to hope that the flavon
couplings to leptons exhibit some kind of nonuniversality,
otherwise the Higgs to tau decays would look completely normal,
with a somewhat higher branching fraction, which in turn may be
attributed to a number of other sources of new physics.
At a muon collider, it will be better to look for the 
$h\rightarrow jj\chi$ modes, which are more abundant
than $h\rightarrow \ell\ell\chi$ anyway. Again, they
will have to be distinguished from
$h\rightarrow\tau\tau\rightarrow jj\nu_\tau\bar{\nu}_\tau$,
which certainly seems possible. First, the missing energy in
the decays of a highly boosted tau is correlated with
the jet direction, and second, hard tau jets are quite
different from light quark jets \cite{tau jets}.

\section{Direct Flavon production}
\label{direct}

In this section we shall consider the production of a real
flavon in association with a real Higgs in both $e^+e^-$ and
$p\bar{p}$ collisions. The resulting signatures will depend
on the fate of the flavon KK states, hence we discuss 
their lifetime first.

The width of a real $\chi_{\bf n}$ decaying back to the
corresponding fermion pair is given by
\beq
\Gamma(\chi_{\bf n}\rightarrow f\bar{f})\ =\
{N_c\over4\pi}\,{y_f^2\over g^2}\, {M_W^2\over M_F^2}\ m_{\bf n}.
\eeq
Depending on $M_F$ (or equivalently,
the size of the extra dimensional space in which
$\chi$ can travel), $\chi_{\bf n}$ can decay promptly at the
interaction vertex, have macroscopic decay length
inside the detector, or decay well outside the detector.
If one assumes that all extra dimensions have
the same size $L$, then the decay length of the KK flavon state
$\chi_{\bf n}$ can be expressed
in terms of $\delta$ and $n$,
\beq
c\tau(\chi_{\bf n}) =
\biggl({M\over 1\ {\rm TeV}}\biggr)^{2(1-{\delta/n})}
\left({1\ {\rm GeV}\over m_{\bf n}}\right)
10^{32{\delta/n}-13} \ m ,
\eeq
and is shown in Table~\ref{decay_length} for various $n$ and $\delta$.
\begin{table}[t!]
\centering
\renewcommand{\arraystretch}{1.2}
\begin{tabular}{||c||c|c|c|c|c|c||}
\hline\hline
& \multicolumn{6}{c||}{$\delta$} \\
\hline
$n$
  & 1
    & 2
      & 3
        & 4
          & 5
            & 6  \\  
\hline\hline
1 & excl.     
    & --- 
      & --- 
        & --- 
          & --- 
            & --- \\
\hline
2 & $10^3$
    & $10^{19}$
      & ---
        & ---
          & ---
            & --- \\
\hline
3 & $5\,10^{-3}$ 
    & $10^8$
      & $10^{19}$
        & ---
          & ---
            & --- \\
\hline
4 & $10^{-5}$ 
    & $10^3$
      & $10^{11}$
        & $10^{19}$
          & ---
            & --- \\
\hline
5 & $10^{-7}$ 
    & $0.6$ 
      & $10^6$
        & $10^{13}$
          & $10^{19}$
            & --- \\
\hline
6 & $10^{-8}$
    & $4\,10^{-3}$ 
      & $10^3$
        & $10^8$
          & $10^{14}$
            & $10^{19}$ \\
\hline\hline
\end{tabular}
\parbox{5.5in}{
\caption{The decay length $c\tau$ in meters of a KK flavon of mass 1 GeV
for $M=1$ TeV and various number of extra dimensions $n$ and $\delta$.
\label{decay_length}}}
\end{table}

For prompt decay, the signal is the same as the virtual $\chi$
exchange which will be discussed in the next section. 
If $\chi_{\bf n}$ has a macroscopic decay length inside the detector,
we will have the spectacular signal of displaced vertices
\cite{Landsberg}. Displaced vertices can appear in several versions
of the supersymmetric extensions of the Standard Model. An example
which gives similar signatures is the low supersymmetry breaking
scale model with neutral wino as the NLSP~\cite{CDM}.
For a range of values for the supersymmetry breaking scale, the NLSP has a
delayed decay to a $Z$ and a gravitino, with $Z$ decaying subsequently
to a pair of leptons or jets. 
Note that the decay length also depends on the mass of the KK mode
and we have a continuous distribution of KK mode masses,
so the range of $M_F$ which can give rise to the displaced vertex
signature is not too narrow.\footnote{In principle the distribution of 
the decay lengths will be different from that of a particle of a single mass.
However, with a low statistics sample,
it will be very difficult to distinguish
these two cases experimentally.}

Finally, if the lifetime of $\chi_{\bf n}$ is 
large enough, it will escape the detector. This will give rise to an
interesting $\rlap{\,/}E$ signature with real Higgs bosons,
somewhat similar to $Zh\rightarrow \nu\bar{\nu}h$.
The difference now is that
the missing mass is not characteristic of the $Z$ mass,
but rather displays a continuous spectrum. If we replace
the Higgs by its vev in the interaction and attach a photon or a gluon, 
we can also get signals
similar to KK graviton production such as $\gamma + \not\!\!\!E$
or $j + \rlap{\,/}E$. Since they have already been studied for the
case of graviton production~\cite{GRW,MPP}
and do not have Higgs associated with it, in the rest of this section
we will instead concentrate on the novel $h + \met$ signals.

At both lepton and hadron colliders the production of
$h\chi_{\bf n}$ will lead predominantly
to a $b\bar{b}\met$ final state for $m_h\lsim 140$ GeV,
and $WW\met$ or $ZZ\met$ for $m_h\gsim 140$ GeV.

The (parton level) cross-section for producing a
single flavon species of mass $m_{\bf n}$ is given by
\beq
\sigma_{\bf n}(\mchi,s)\ \equiv\ 
{1\over \rho(m_{\bf n})}
{d\sigma\over dm^2_{\bf n}}
\ =\ {1\over 64\pi}\,{y_f^2\over M_F^2}\
\lambda^{1\over2}(1,{m_h^2/s},{m_{\bf n}^2/s}),
\label{xsec mn}
\eeq
where
\beq
\lambda(x,y,z)\ \equiv\ x^2+y^2+z^2-2xy-2xz-2yz
\eeq
and $\sqrt{s}$ is the center of mass energy of the
collision. To get the total cross-section $\sigma_{tot}(s)$,
one needs to integrate over the available mass range
for the KK states
$\{m_0,\sqrt{s}-m_h\}$:
\bear
\sigma_{tot}(s)&=&\int_{m_0}^{\sqrt{s}-m_h} 
d\mchi\,2\mchi\,\rho(m_{\bf n})\,\sigma_{\bf n}(\mchi,s)
\label{xsec real emission} \\
&=&{y_f^2\over 64\pi M^{\delta+2}}\,
{1\over(4\pi)^{\delta/2}\Gamma({\delta\over2})}
\ \int_{m_0}^{\sqrt{s}-m_h}
d\mchi\, 2\mchi\, (m^2_{\bf n}-m_0^2)^{(\delta-2)\over2}\,
\lambda^{1\over2}(1,{m_h^2\over s},{m_{\bf n}^2\over s}).
\nonumber
\eear

In the remaining of this section we shall discuss
the case of lepton and hadron colliders separately.
In the latter case, the cross-section (\ref{xsec real emission})
in addition has to be integrated over the respective
parton distribution functions $f^{p}(x)$, e.g. at the Tevatron
we have
\beq
\sigma_{tot}(E_{\rm CM})\ = \
\int_0^1 dx_1 \int_0^1 dx_2 \int_{m_0}^{\sqrt{x_1x_2}E_{\rm CM}-m_h} dm_{\bf n}\,
2m_{\bf n} \rho(m_{\bf n}) f^{p}(x_1) f^{\bar{p}}(x_2) 
\sigma_{\bf n}(\mchi,x_1x_2E^2_{\rm CM})
\label{hadronic xsec}
\eeq
with $E_{\rm CM}=2$ TeV now being the total center of mass beam energy.

\subsection{Lepton colliders}
\label{direct-lepton colliders}

\begin{figure}[t!]
\centerline{\psfig{figure=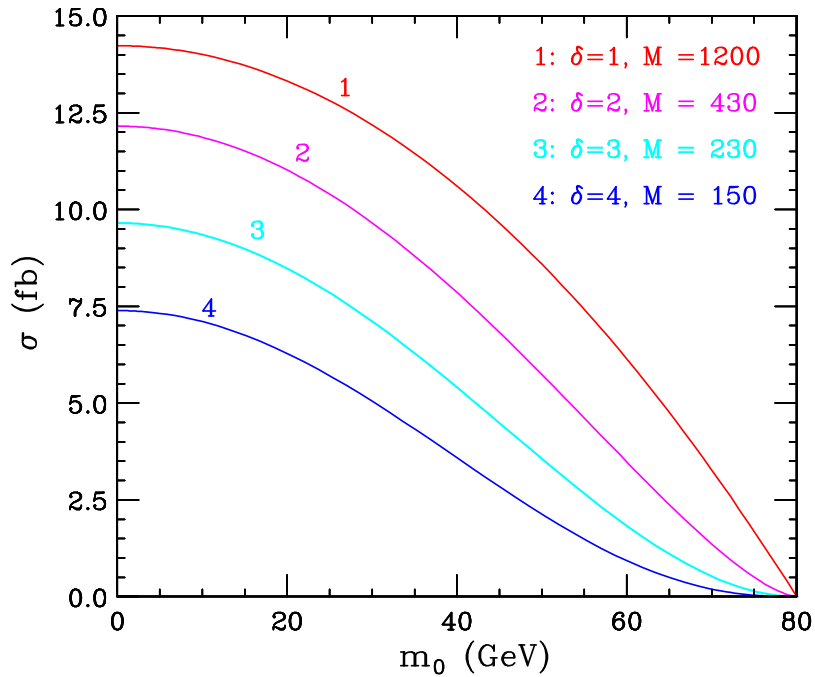,height=3.5in}}
\begin{center}
\parbox{5.5in}{
\caption[] {\small Signal cross-section for $h\chi$ production at LEP
with $\sqrt{s}=200$ GeV, for $m_h=120$ GeV and different values of
$\delta$ and $M$.
\label{chiH_LEP}}}
\end{center}
\end{figure}

For $m_h+m_0\lsim 200$ GeV, KK modes of the flavon can
in principle be produced at LEP. The total cross-section
(\ref{xsec real emission}) for $e^+e^-\rightarrow h\chi$ 
with $\sqrt{s}=200$ GeV is shown in Fig.~\ref{chiH_LEP},
for $m_h=120$ GeV and several values of $\delta$ and $M$.
For a different $M$, the cross-section can be easily obtained
from the scaling relation $\sigma \propto M^{-(\delta+2)}$.
The cross-section maxes out at small values of 
$m_0$, when there are more KK states kinematically available.
From Fig.~\ref{chiH_LEP} we see that if $\delta=n$, 
the typical maximal values of the 
cross-section are around 10 fb,
in which case only one or two events can be
expected at LEP. Of course, the bounds on $M$ from gravity are weaker
if one considers cases of $\delta<n$,
then LEP might be able to explore the region of very low values of $M$
and small $\delta$. 
Given the current limits on the Higgs mass,
one obviously has to go to a higher energy machine in order
to start to see an appreciable signal. We therefore turn to
discuss the NLC case in a little bit more detail next.

\begin{figure}[t!]
\centerline{\psfig{figure=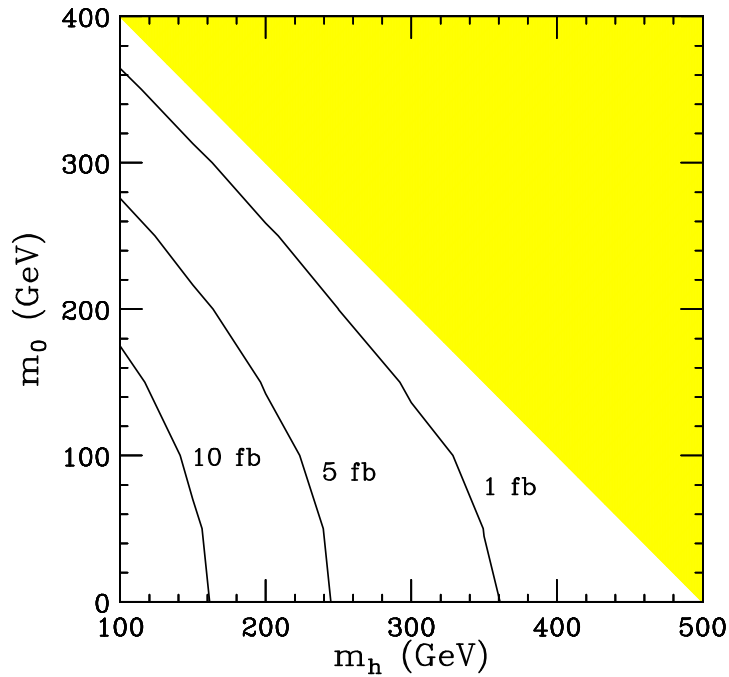,height=3.5in}}
\begin{center}
\parbox{5.5in}{
\caption[] {\small Contour plot of the total cross-section for real
flavon emission at the NLC with $\sqrt{s}=500$ GeV, for
$y_f=1$, $\delta=2$ and $M=1$ TeV. The shaded area is not
kinematically accessible at this center-of-mass energy.
\label{xsec real NLC}}}
\end{center}
\end{figure}

In Fig.~\ref{xsec real NLC} we show a contour plot of the
total integrated cross-section at an NLC with $\sqrt{s}=500$ GeV.
We fix $y_f=1$, $\delta=2$, $M=1$ TeV and vary
the Higgs mass $m_h$ and the mass $m_0$ of the flavon zero mode.
The cross-section shown in Fig.~\ref{xsec real NLC}
can be easily rescaled for different values of $y_f$ or $M$,
recall that it scales like $y_f^2/M^{2+\delta}$.
With higher energy and larger integrated luminosity, the flavon
production cross-section at the NLC can provide a much
more statistically significant signal sample.
With 50 ${\rm fb}^{-1}$ of data for the parameters
in Fig.~\ref{xsec real NLC}, one will have a few
hundred signal events available. We will then focus our
discussion on the question of cuts optimization and
increasing the signal-to-background ratio.

The search strategy at the NLC will depend
on the Higgs mass. For heavy ($\gsim140$ GeV) Higgs masses,
the signal will be $WW\not\!\!\!E$, similar to chargino
pair production with charginos decaying to real $W$'s and
the lightest neutralino. In that case, it is best to
reconstruct the $W$'s from the four jet final state
\cite{JLC}, where the only non-negligible background
is $WWZ$ with $Z\rightarrow\nu\bar{\nu}$, which is on the order of
10 fb \cite{Hitoshi}, after taking into account the
$Z$ branching ratio into neutrinos. With cleverly
designed cuts \cite{more cuts}, it is easy to see
an excess over the background for a signal of ${\cal O}$(1 fb),
which can then be used to put constraints on $m_0$, $m_h$, and $M$.

\begin{figure}[t!]
\begin{center}
{
\unitlength=0.8 pt
\SetScale{0.8}
\SetWidth{0.7}      
\scriptsize    
{} \qquad\allowbreak
\begin{picture}(95,99)(0,0)
\Text(15.0,80.0)[r]{$e$}
\ArrowLine(16.0,80.0)(37.0,80.0) 
\Text(47.0,81.0)[b]{$\gamma$}
\DashLine(37.0,80.0)(58.0,80.0){3.0} 
\Text(80.0,90.0)[l]{$b$}
\ArrowLine(58.0,80.0)(79.0,90.0) 
\Text(80.0,70.0)[l]{$\bar{b}$}
\ArrowLine(79.0,70.0)(58.0,80.0) 
\Text(33.0,70.0)[r]{$e$}
\ArrowLine(37.0,80.0)(37.0,60.0) 
\Line(37.0,60.0)(58.0,60.0) 
\Text(80.0,50.0)[l]{$\nu_e$}
\ArrowLine(58.0,60.0)(79.0,50.0) 
\Text(33.0,50.0)[r]{$W^+$}
\DashArrowLine(37.0,40.0)(37.0,60.0){3.0} 
\Text(15.0,40.0)[r]{$\bar{e}$}
\ArrowLine(37.0,40.0)(16.0,40.0) 
\Line(37.0,40.0)(58.0,40.0) 
\Text(80.0,30.0)[l]{$\bar{\nu}_e$}
\ArrowLine(79.0,30.0)(58.0,40.0) 
\Text(47,0)[b] {diagr. 1}
\end{picture} \ 
{} \qquad\allowbreak
\begin{picture}(95,99)(0,0)
\Text(15.0,80.0)[r]{$e$}
\ArrowLine(16.0,80.0)(37.0,80.0) 
\Text(47.0,81.0)[b]{$\gamma$}
\DashLine(37.0,80.0)(58.0,80.0){3.0} 
\Text(80.0,90.0)[l]{$b$}
\ArrowLine(58.0,80.0)(79.0,90.0) 
\Text(80.0,70.0)[l]{$\bar{b}$}
\ArrowLine(79.0,70.0)(58.0,80.0) 
\Text(33.0,60.0)[r]{$e$}
\ArrowLine(37.0,80.0)(37.0,40.0) 
\Text(15.0,40.0)[r]{$\bar{e}$}
\ArrowLine(37.0,40.0)(16.0,40.0) 
\Text(47.0,41.0)[b]{$Z$}
\DashLine(37.0,40.0)(58.0,40.0){3.0} 
\Text(80.0,50.0)[l]{$\nu_e$}
\ArrowLine(58.0,40.0)(79.0,50.0) 
\Text(80.0,30.0)[l]{$\bar{\nu}_e$}
\ArrowLine(79.0,30.0)(58.0,40.0) 
\Text(47,0)[b] {diagr.2}
\end{picture} \ 
{} \qquad\allowbreak
\begin{picture}(95,99)(0,0)
\Text(15.0,90.0)[r]{$e$}
\ArrowLine(16.0,90.0)(37.0,80.0) 
\Text(15.0,70.0)[r]{$\bar{e}$}
\ArrowLine(37.0,80.0)(16.0,70.0) 
\Text(47.0,81.0)[b]{$\gamma$}
\DashLine(37.0,80.0)(58.0,80.0){3.0} 
\Text(80.0,90.0)[l]{$\bar{b}$}
\ArrowLine(79.0,90.0)(58.0,80.0) 
\Text(54.0,70.0)[r]{$b$}
\ArrowLine(58.0,80.0)(58.0,60.0) 
\Text(80.0,70.0)[l]{$b$}
\ArrowLine(58.0,60.0)(79.0,70.0) 
\Text(57.0,50.0)[r]{$Z$}
\DashLine(58.0,60.0)(58.0,40.0){3.0} 
\Text(80.0,50.0)[l]{$\nu_e$}
\ArrowLine(58.0,40.0)(79.0,50.0) 
\Text(80.0,30.0)[l]{$\bar{\nu}_e$}
\ArrowLine(79.0,30.0)(58.0,40.0) 
\Text(47,0)[b] {diagr.3}
\end{picture} \ 
{} \qquad\allowbreak
\begin{picture}(95,99)(0,0)
\Text(15.0,90.0)[r]{$e$}
\ArrowLine(16.0,90.0)(37.0,80.0) 
\Text(15.0,70.0)[r]{$\bar{e}$}
\ArrowLine(37.0,80.0)(16.0,70.0) 
\Text(47.0,81.0)[b]{$\gamma$}
\DashLine(37.0,80.0)(58.0,80.0){3.0} 
\Text(80.0,90.0)[l]{$b$}
\ArrowLine(58.0,80.0)(79.0,90.0) 
\Text(54.0,70.0)[r]{$b$}
\ArrowLine(58.0,60.0)(58.0,80.0) 
\Text(80.0,70.0)[l]{$\bar{b}$}
\ArrowLine(79.0,70.0)(58.0,60.0) 
\Text(57.0,50.0)[r]{$Z$}
\DashLine(58.0,60.0)(58.0,40.0){3.0} 
\Text(80.0,50.0)[l]{$\nu_e$}
\ArrowLine(58.0,40.0)(79.0,50.0) 
\Text(80.0,30.0)[l]{$\bar{\nu}_e$}
\ArrowLine(79.0,30.0)(58.0,40.0) 
\Text(47,0)[b] {diagr.4}
\end{picture} \ 
{} \qquad\allowbreak
\begin{picture}(95,99)(0,0)
\Text(15.0,80.0)[r]{$e$}
\ArrowLine(16.0,80.0)(37.0,80.0) 
\Line(37.0,80.0)(58.0,80.0) 
\Text(80.0,90.0)[l]{$\nu_e$}
\ArrowLine(58.0,80.0)(79.0,90.0) 
\Text(33.0,70.0)[r]{$W^+$}
\DashArrowLine(37.0,60.0)(37.0,80.0){3.0} 
\Text(15.0,60.0)[r]{$\bar{e}$}
\ArrowLine(37.0,60.0)(16.0,60.0) 
\Text(47.0,64.0)[b]{$\nu_e$}
\ArrowLine(58.0,60.0)(37.0,60.0) 
\Text(80.0,70.0)[l]{$\bar{\nu}_e$}
\ArrowLine(79.0,70.0)(58.0,60.0) 
\Text(57.0,50.0)[r]{$Z$}
\DashLine(58.0,60.0)(58.0,40.0){3.0} 
\Text(80.0,50.0)[l]{$b$}
\ArrowLine(58.0,40.0)(79.0,50.0) 
\Text(80.0,30.0)[l]{$\bar{b}$}
\ArrowLine(79.0,30.0)(58.0,40.0) 
\Text(47,0)[b] {diagr.5}
\end{picture} \ 
{} \qquad\allowbreak
\begin{picture}(95,99)(0,0)
\Text(15.0,80.0)[r]{$e$}
\ArrowLine(16.0,80.0)(37.0,80.0) 
\Line(37.0,80.0)(58.0,80.0) 
\Text(80.0,90.0)[l]{$\nu_e$}
\ArrowLine(58.0,80.0)(79.0,90.0) 
\Text(33.0,70.0)[r]{$W^+$}
\DashArrowLine(37.0,60.0)(37.0,80.0){3.0} 
\Line(37.0,60.0)(58.0,60.0) 
\Text(80.0,70.0)[l]{$\bar{\nu}_e$}
\ArrowLine(79.0,70.0)(58.0,60.0) 
\Text(33.0,50.0)[r]{$e$}
\ArrowLine(37.0,60.0)(37.0,40.0) 
\Text(15.0,40.0)[r]{$\bar{e}$}
\ArrowLine(37.0,40.0)(16.0,40.0) 
\Text(47.0,41.0)[b]{$\gamma$}
\DashLine(37.0,40.0)(58.0,40.0){3.0} 
\Text(80.0,50.0)[l]{$b$}
\ArrowLine(58.0,40.0)(79.0,50.0) 
\Text(80.0,30.0)[l]{$\bar{b}$}
\ArrowLine(79.0,30.0)(58.0,40.0) 
\Text(47,0)[b] {diagr.6}
\end{picture} \ 
{} \qquad\allowbreak
\begin{picture}(95,99)(0,0)
\Text(15.0,80.0)[r]{$e$}
\ArrowLine(16.0,80.0)(37.0,80.0) 
\Line(37.0,80.0)(58.0,80.0) 
\Text(80.0,90.0)[l]{$\nu_e$}
\ArrowLine(58.0,80.0)(79.0,90.0) 
\Text(33.0,70.0)[r]{$W^+$}
\DashArrowLine(37.0,60.0)(37.0,80.0){3.0} 
\Line(37.0,60.0)(58.0,60.0) 
\Text(80.0,70.0)[l]{$\bar{\nu}_e$}
\ArrowLine(79.0,70.0)(58.0,60.0) 
\Text(33.0,50.0)[r]{$e$}
\ArrowLine(37.0,60.0)(37.0,40.0) 
\Text(15.0,40.0)[r]{$\bar{e}$}
\ArrowLine(37.0,40.0)(16.0,40.0) 
\Text(47.0,41.0)[b]{$Z$}
\DashLine(37.0,40.0)(58.0,40.0){3.0} 
\Text(80.0,50.0)[l]{$b$}
\ArrowLine(58.0,40.0)(79.0,50.0) 
\Text(80.0,30.0)[l]{$\bar{b}$}
\ArrowLine(79.0,30.0)(58.0,40.0) 
\Text(47,0)[b] {diagr.7}
\end{picture} \ 
{} \qquad\allowbreak
\begin{picture}(95,99)(0,0)
\Text(15.0,90.0)[r]{$e$}
\ArrowLine(16.0,90.0)(58.0,90.0) 
\Text(80.0,90.0)[l]{$\nu_e$}
\ArrowLine(58.0,90.0)(79.0,90.0) 
\Text(54.0,80.0)[r]{$W^+$}
\DashArrowLine(58.0,70.0)(58.0,90.0){3.0} 
\Text(80.0,70.0)[l]{$b$}
\ArrowLine(58.0,70.0)(79.0,70.0) 
\Text(54.0,60.0)[r]{$u$}
\ArrowLine(58.0,50.0)(58.0,70.0) 
\Text(80.0,50.0)[l]{$\bar{b}$}
\ArrowLine(79.0,50.0)(58.0,50.0) 
\Text(54.0,40.0)[r]{$W^+$}
\DashArrowLine(58.0,30.0)(58.0,50.0){3.0} 
\Text(15.0,30.0)[r]{$\bar{e}$}
\ArrowLine(58.0,30.0)(16.0,30.0) 
\Text(80.0,30.0)[l]{$\bar{\nu}_e$}
\ArrowLine(79.0,30.0)(58.0,30.0) 
\Text(47,0)[b] {diagr.8}
\end{picture} \ 
{} \qquad\allowbreak
\begin{picture}(95,99)(0,0)
\Text(15.0,90.0)[r]{$e$}
\ArrowLine(16.0,90.0)(58.0,90.0) 
\Text(80.0,90.0)[l]{$\nu_e$}
\ArrowLine(58.0,90.0)(79.0,90.0) 
\Text(54.0,80.0)[r]{$W^+$}
\DashArrowLine(58.0,70.0)(58.0,90.0){3.0} 
\Text(80.0,70.0)[l]{$b$}
\ArrowLine(58.0,70.0)(79.0,70.0) 
\Text(54.0,60.0)[r]{$c$}
\ArrowLine(58.0,50.0)(58.0,70.0) 
\Text(80.0,50.0)[l]{$\bar{b}$}
\ArrowLine(79.0,50.0)(58.0,50.0) 
\Text(54.0,40.0)[r]{$W^+$}
\DashArrowLine(58.0,30.0)(58.0,50.0){3.0} 
\Text(15.0,30.0)[r]{$\bar{e}$}
\ArrowLine(58.0,30.0)(16.0,30.0) 
\Text(80.0,30.0)[l]{$\bar{\nu}_e$}
\ArrowLine(79.0,30.0)(58.0,30.0) 
\Text(47,0)[b] {diagr.9}
\end{picture} \ 
{} \qquad\allowbreak
\begin{picture}(95,99)(0,0)
\Text(15.0,90.0)[r]{$e$}
\ArrowLine(16.0,90.0)(58.0,90.0) 
\Text(80.0,90.0)[l]{$\nu_e$}
\ArrowLine(58.0,90.0)(79.0,90.0) 
\Text(54.0,80.0)[r]{$W^+$}
\DashArrowLine(58.0,70.0)(58.0,90.0){3.0} 
\Text(80.0,70.0)[l]{$b$}
\ArrowLine(58.0,70.0)(79.0,70.0) 
\Text(54.0,60.0)[r]{$t$}
\ArrowLine(58.0,50.0)(58.0,70.0) 
\Text(80.0,50.0)[l]{$\bar{b}$}
\ArrowLine(79.0,50.0)(58.0,50.0) 
\Text(54.0,40.0)[r]{$W^+$}
\DashArrowLine(58.0,30.0)(58.0,50.0){3.0} 
\Text(15.0,30.0)[r]{$\bar{e}$}
\ArrowLine(58.0,30.0)(16.0,30.0) 
\Text(80.0,30.0)[l]{$\bar{\nu}_e$}
\ArrowLine(79.0,30.0)(58.0,30.0) 
\Text(47,0)[b] {diagr.10}
\end{picture} \ 
{} \qquad\allowbreak
\begin{picture}(95,99)(0,0)
\Text(15.0,80.0)[r]{$e$}
\ArrowLine(16.0,80.0)(37.0,80.0) 
\Line(37.0,80.0)(58.0,80.0) 
\Text(80.0,90.0)[l]{$\nu_e$}
\ArrowLine(58.0,80.0)(79.0,90.0) 
\Text(33.0,70.0)[r]{$W^+$}
\DashArrowLine(37.0,60.0)(37.0,80.0){3.0} 
\Text(47.0,61.0)[b]{$\gamma$}
\DashLine(37.0,60.0)(58.0,60.0){3.0} 
\Text(80.0,70.0)[l]{$b$}
\ArrowLine(58.0,60.0)(79.0,70.0) 
\Text(80.0,50.0)[l]{$\bar{b}$}
\ArrowLine(79.0,50.0)(58.0,60.0) 
\Text(33.0,50.0)[r]{$W^+$}
\DashArrowLine(37.0,40.0)(37.0,60.0){3.0} 
\Text(15.0,40.0)[r]{$\bar{e}$}
\ArrowLine(37.0,40.0)(16.0,40.0) 
\Line(37.0,40.0)(58.0,40.0) 
\Text(80.0,30.0)[l]{$\bar{\nu}_e$}
\ArrowLine(79.0,30.0)(58.0,40.0) 
\Text(47,0)[b] {diagr.11}
\end{picture} \ 
{} \qquad\allowbreak
\begin{picture}(95,99)(0,0)
\Text(15.0,80.0)[r]{$e$}
\ArrowLine(16.0,80.0)(37.0,80.0) 
\Line(37.0,80.0)(58.0,80.0) 
\Text(80.0,90.0)[l]{$\nu_e$}
\ArrowLine(58.0,80.0)(79.0,90.0) 
\Text(33.0,70.0)[r]{$W^+$}
\DashArrowLine(37.0,60.0)(37.0,80.0){3.0} 
\Text(47.0,61.0)[b]{$H$}
\DashLine(37.0,60.0)(58.0,60.0){1.0}
\Text(80.0,70.0)[l]{$b$}
\ArrowLine(58.0,60.0)(79.0,70.0) 
\Text(80.0,50.0)[l]{$\bar{b}$}
\ArrowLine(79.0,50.0)(58.0,60.0) 
\Text(33.0,50.0)[r]{$W^+$}
\DashArrowLine(37.0,40.0)(37.0,60.0){3.0} 
\Text(15.0,40.0)[r]{$\bar{e}$}
\ArrowLine(37.0,40.0)(16.0,40.0) 
\Line(37.0,40.0)(58.0,40.0) 
\Text(80.0,30.0)[l]{$\bar{\nu}_e$}
\ArrowLine(79.0,30.0)(58.0,40.0) 
\Text(47,0)[b] {diagr.12}
\end{picture} \ 
{} \qquad\allowbreak
\begin{picture}(95,99)(0,0)
\Text(15.0,80.0)[r]{$e$}
\ArrowLine(16.0,80.0)(37.0,80.0) 
\Line(37.0,80.0)(58.0,80.0) 
\Text(80.0,90.0)[l]{$\nu_e$}
\ArrowLine(58.0,80.0)(79.0,90.0) 
\Text(33.0,70.0)[r]{$W^+$}
\DashArrowLine(37.0,60.0)(37.0,80.0){3.0} 
\Text(47.0,61.0)[b]{$Z$}
\DashLine(37.0,60.0)(58.0,60.0){3.0} 
\Text(80.0,70.0)[l]{$b$}
\ArrowLine(58.0,60.0)(79.0,70.0) 
\Text(80.0,50.0)[l]{$\bar{b}$}
\ArrowLine(79.0,50.0)(58.0,60.0) 
\Text(33.0,50.0)[r]{$W^+$}
\DashArrowLine(37.0,40.0)(37.0,60.0){3.0} 
\Text(15.0,40.0)[r]{$\bar{e}$}
\ArrowLine(37.0,40.0)(16.0,40.0) 
\Line(37.0,40.0)(58.0,40.0) 
\Text(80.0,30.0)[l]{$\bar{\nu}_e$}
\ArrowLine(79.0,30.0)(58.0,40.0) 
\Text(47,0)[b] {diagr.13}
\end{picture} \ 
{} \qquad\allowbreak
\begin{picture}(95,99)(0,0)
\Text(15.0,80.0)[r]{$e$}
\ArrowLine(16.0,80.0)(37.0,80.0) 
\Text(47.0,84.0)[b]{$\nu_e$}
\ArrowLine(37.0,80.0)(58.0,80.0) 
\Text(80.0,90.0)[l]{$\nu_e$}
\ArrowLine(58.0,80.0)(79.0,90.0) 
\Text(57.0,70.0)[r]{$Z$}
\DashLine(58.0,80.0)(58.0,60.0){3.0} 
\Text(80.0,70.0)[l]{$b$}
\ArrowLine(58.0,60.0)(79.0,70.0) 
\Text(80.0,50.0)[l]{$\bar{b}$}
\ArrowLine(79.0,50.0)(58.0,60.0) 
\Text(33.0,60.0)[r]{$W^+$}
\DashArrowLine(37.0,40.0)(37.0,80.0){3.0} 
\Text(15.0,40.0)[r]{$\bar{e}$}
\ArrowLine(37.0,40.0)(16.0,40.0) 
\Line(37.0,40.0)(58.0,40.0) 
\Text(80.0,30.0)[l]{$\bar{\nu}_e$}
\ArrowLine(79.0,30.0)(58.0,40.0) 
\Text(47,0)[b] {diagr.14}
\end{picture} \ 
{} \qquad\allowbreak
\begin{picture}(95,99)(0,0)
\Text(15.0,90.0)[r]{$e$}
\ArrowLine(16.0,90.0)(37.0,80.0) 
\Text(15.0,70.0)[r]{$\bar{e}$}
\ArrowLine(37.0,80.0)(16.0,70.0) 
\Text(47.0,81.0)[b]{$Z$}
\DashLine(37.0,80.0)(58.0,80.0){3.0} 
\Text(80.0,90.0)[l]{$\bar{\nu}_e$}
\ArrowLine(79.0,90.0)(58.0,80.0) 
\Text(54.0,70.0)[r]{$\nu_e$}
\ArrowLine(58.0,80.0)(58.0,60.0) 
\Text(80.0,70.0)[l]{$\nu_e$}
\ArrowLine(58.0,60.0)(79.0,70.0) 
\Text(57.0,50.0)[r]{$Z$}
\DashLine(58.0,60.0)(58.0,40.0){3.0} 
\Text(80.0,50.0)[l]{$b$}
\ArrowLine(58.0,40.0)(79.0,50.0) 
\Text(80.0,30.0)[l]{$\bar{b}$}
\ArrowLine(79.0,30.0)(58.0,40.0) 
\Text(47,0)[b] {diagr.15}
\end{picture} \ 
{} \qquad\allowbreak
\begin{picture}(95,99)(0,0)
\Text(15.0,90.0)[r]{$e$}
\ArrowLine(16.0,90.0)(37.0,80.0) 
\Text(15.0,70.0)[r]{$\bar{e}$}
\ArrowLine(37.0,80.0)(16.0,70.0) 
\Text(47.0,81.0)[b]{$Z$}
\DashLine(37.0,80.0)(58.0,80.0){3.0} 
\Text(80.0,90.0)[l]{$\nu_e$}
\ArrowLine(58.0,80.0)(79.0,90.0) 
\Text(54.0,70.0)[r]{$\nu_e$}
\ArrowLine(58.0,60.0)(58.0,80.0) 
\Text(80.0,70.0)[l]{$\bar{\nu}_e$}
\ArrowLine(79.0,70.0)(58.0,60.0) 
\Text(57.0,50.0)[r]{$Z$}
\DashLine(58.0,60.0)(58.0,40.0){3.0} 
\Text(80.0,50.0)[l]{$b$}
\ArrowLine(58.0,40.0)(79.0,50.0) 
\Text(80.0,30.0)[l]{$\bar{b}$}
\ArrowLine(79.0,30.0)(58.0,40.0) 
\Text(47,0)[b] {diagr.16}
\end{picture} \ 
{} \qquad\allowbreak
\begin{picture}(95,99)(0,0)
\Text(15.0,90.0)[r]{$e$}
\ArrowLine(16.0,90.0)(37.0,80.0) 
\Text(15.0,70.0)[r]{$\bar{e}$}
\ArrowLine(37.0,80.0)(16.0,70.0) 
\Text(47.0,81.0)[b]{$Z$}
\DashLine(37.0,80.0)(58.0,80.0){3.0} 
\Text(80.0,90.0)[l]{$\bar{b}$}
\ArrowLine(79.0,90.0)(58.0,80.0) 
\Text(54.0,70.0)[r]{$b$}
\ArrowLine(58.0,80.0)(58.0,60.0) 
\Text(80.0,70.0)[l]{$b$}
\ArrowLine(58.0,60.0)(79.0,70.0) 
\Text(57.0,50.0)[r]{$Z$}
\DashLine(58.0,60.0)(58.0,40.0){3.0} 
\Text(80.0,50.0)[l]{$\nu_e$}
\ArrowLine(58.0,40.0)(79.0,50.0) 
\Text(80.0,30.0)[l]{$\bar{\nu}_e$}
\ArrowLine(79.0,30.0)(58.0,40.0) 
\Text(47,0)[b] {diagr.17}
\end{picture} \ 
{} \qquad\allowbreak
\begin{picture}(95,99)(0,0)
\Text(15.0,90.0)[r]{$e$}
\ArrowLine(16.0,90.0)(37.0,80.0) 
\Text(15.0,70.0)[r]{$\bar{e}$}
\ArrowLine(37.0,80.0)(16.0,70.0) 
\Text(47.0,81.0)[b]{$Z$}
\DashLine(37.0,80.0)(58.0,80.0){3.0} 
\Text(80.0,90.0)[l]{$b$}
\ArrowLine(58.0,80.0)(79.0,90.0) 
\Text(54.0,70.0)[r]{$b$}
\ArrowLine(58.0,60.0)(58.0,80.0) 
\Text(80.0,70.0)[l]{$\bar{b}$}
\ArrowLine(79.0,70.0)(58.0,60.0) 
\Text(57.0,50.0)[r]{$Z$}
\DashLine(58.0,60.0)(58.0,40.0){3.0} 
\Text(80.0,50.0)[l]{$\nu_e$}
\ArrowLine(58.0,40.0)(79.0,50.0) 
\Text(80.0,30.0)[l]{$\bar{\nu}_e$}
\ArrowLine(79.0,30.0)(58.0,40.0) 
\Text(47,0)[b] {diagr.18}
\end{picture} \ 
{} \qquad\allowbreak
\begin{picture}(95,99)(0,0)
\Text(15.0,70.0)[r]{$e$}
\ArrowLine(16.0,70.0)(37.0,60.0) 
\Text(15.0,50.0)[r]{$\bar{e}$}
\ArrowLine(37.0,60.0)(16.0,50.0) 
\Text(37.0,62.0)[lb]{$Z$}
\DashLine(37.0,60.0)(58.0,60.0){3.0} 
\Text(57.0,70.0)[r]{$H$}
\DashLine(58.0,60.0)(58.0,80.0){1.0}
\Text(80.0,90.0)[l]{$b$}
\ArrowLine(58.0,80.0)(79.0,90.0) 
\Text(80.0,70.0)[l]{$\bar{b}$}
\ArrowLine(79.0,70.0)(58.0,80.0) 
\Text(57.0,50.0)[r]{$Z$}
\DashLine(58.0,60.0)(58.0,40.0){3.0} 
\Text(80.0,50.0)[l]{$\nu_e$}
\ArrowLine(58.0,40.0)(79.0,50.0) 
\Text(80.0,30.0)[l]{$\bar{\nu}_e$}
\ArrowLine(79.0,30.0)(58.0,40.0) 
\Text(47,0)[b] {diagr.19}
\end{picture} \ 
{} \qquad\allowbreak
\begin{picture}(95,99)(0,0)
\Text(15.0,80.0)[r]{$e$}
\ArrowLine(16.0,80.0)(37.0,80.0) 
\Text(47.0,81.0)[b]{$Z$}
\DashLine(37.0,80.0)(58.0,80.0){3.0} 
\Text(80.0,90.0)[l]{$\nu_e$}
\ArrowLine(58.0,80.0)(79.0,90.0) 
\Text(80.0,70.0)[l]{$\bar{\nu}_e$}
\ArrowLine(79.0,70.0)(58.0,80.0) 
\Text(33.0,60.0)[r]{$e$}
\ArrowLine(37.0,80.0)(37.0,40.0) 
\Text(15.0,40.0)[r]{$\bar{e}$}
\ArrowLine(37.0,40.0)(16.0,40.0) 
\Text(47.0,41.0)[b]{$\gamma$}
\DashLine(37.0,40.0)(58.0,40.0){3.0} 
\Text(80.0,50.0)[l]{$b$}
\ArrowLine(58.0,40.0)(79.0,50.0) 
\Text(80.0,30.0)[l]{$\bar{b}$}
\ArrowLine(79.0,30.0)(58.0,40.0) 
\Text(47,0)[b] {diagr.20}
\end{picture} \ 
{} \qquad\allowbreak
\begin{picture}(95,99)(0,0)
\Text(15.0,80.0)[r]{$e$}
\ArrowLine(16.0,80.0)(37.0,80.0) 
\Text(47.0,81.0)[b]{$Z$}
\DashLine(37.0,80.0)(58.0,80.0){3.0} 
\Text(80.0,90.0)[l]{$b$}
\ArrowLine(58.0,80.0)(79.0,90.0) 
\Text(80.0,70.0)[l]{$\bar{b}$}
\ArrowLine(79.0,70.0)(58.0,80.0) 
\Text(33.0,70.0)[r]{$e$}
\ArrowLine(37.0,80.0)(37.0,60.0) 
\Line(37.0,60.0)(58.0,60.0) 
\Text(80.0,50.0)[l]{$\nu_e$}
\ArrowLine(58.0,60.0)(79.0,50.0) 
\Text(33.0,50.0)[r]{$W^+$}
\DashArrowLine(37.0,40.0)(37.0,60.0){3.0} 
\Text(15.0,40.0)[r]{$\bar{e}$}
\ArrowLine(37.0,40.0)(16.0,40.0) 
\Line(37.0,40.0)(58.0,40.0) 
\Text(80.0,30.0)[l]{$\bar{\nu}_e$}
\ArrowLine(79.0,30.0)(58.0,40.0) 
\Text(47,0)[b] {diagr.21}
\end{picture} \ 
{} \qquad\allowbreak
\begin{picture}(95,99)(0,0)
\Text(15.0,80.0)[r]{$e$}
\ArrowLine(16.0,80.0)(37.0,80.0) 
\Text(47.0,81.0)[b]{$Z$}
\DashLine(37.0,80.0)(58.0,80.0){3.0} 
\Text(80.0,90.0)[l]{$b$}
\ArrowLine(58.0,80.0)(79.0,90.0) 
\Text(80.0,70.0)[l]{$\bar{b}$}
\ArrowLine(79.0,70.0)(58.0,80.0) 
\Text(33.0,60.0)[r]{$e$}
\ArrowLine(37.0,80.0)(37.0,40.0) 
\Text(15.0,40.0)[r]{$\bar{e}$}
\ArrowLine(37.0,40.0)(16.0,40.0) 
\Text(47.0,41.0)[b]{$Z$}
\DashLine(37.0,40.0)(58.0,40.0){3.0} 
\Text(80.0,50.0)[l]{$\nu_e$}
\ArrowLine(58.0,40.0)(79.0,50.0) 
\Text(80.0,30.0)[l]{$\bar{\nu}_e$}
\ArrowLine(79.0,30.0)(58.0,40.0) 
\Text(47,0)[b] {diagr.22}
\end{picture} \ 
{} \qquad\allowbreak
\begin{picture}(95,99)(0,0)
\Text(15.0,80.0)[r]{$e$}
\ArrowLine(16.0,80.0)(37.0,80.0) 
\Text(47.0,81.0)[b]{$Z$}
\DashLine(37.0,80.0)(58.0,80.0){3.0} 
\Text(80.0,90.0)[l]{$\nu_e$}
\ArrowLine(58.0,80.0)(79.0,90.0) 
\Text(80.0,70.0)[l]{$\bar{\nu}_e$}
\ArrowLine(79.0,70.0)(58.0,80.0) 
\Text(33.0,60.0)[r]{$e$}
\ArrowLine(37.0,80.0)(37.0,40.0) 
\Text(15.0,40.0)[r]{$\bar{e}$}
\ArrowLine(37.0,40.0)(16.0,40.0) 
\Text(47.0,41.0)[b]{$Z$}
\DashLine(37.0,40.0)(58.0,40.0){3.0} 
\Text(80.0,50.0)[l]{$b$}
\ArrowLine(58.0,40.0)(79.0,50.0) 
\Text(80.0,30.0)[l]{$\bar{b}$}
\ArrowLine(79.0,30.0)(58.0,40.0) 
\Text(47,0)[b] {diagr.23}
\end{picture} \ 
}
\parbox{5.5in}{
\caption[] {\small The diagrams for the process
$e^+e^-\rightarrow b\bar{b}\nu_e\bar{\nu}_e$. The subset without
$W$ bosons applies for  
$e^+e^-\rightarrow b\bar{b}\nu_\mu\bar{\nu}_\mu$ and
$e^+e^-\rightarrow b\bar{b}\nu_\tau\bar{\nu}_\tau$.
\label{diagrams_NNBB}}}
\end{center}
\end{figure}
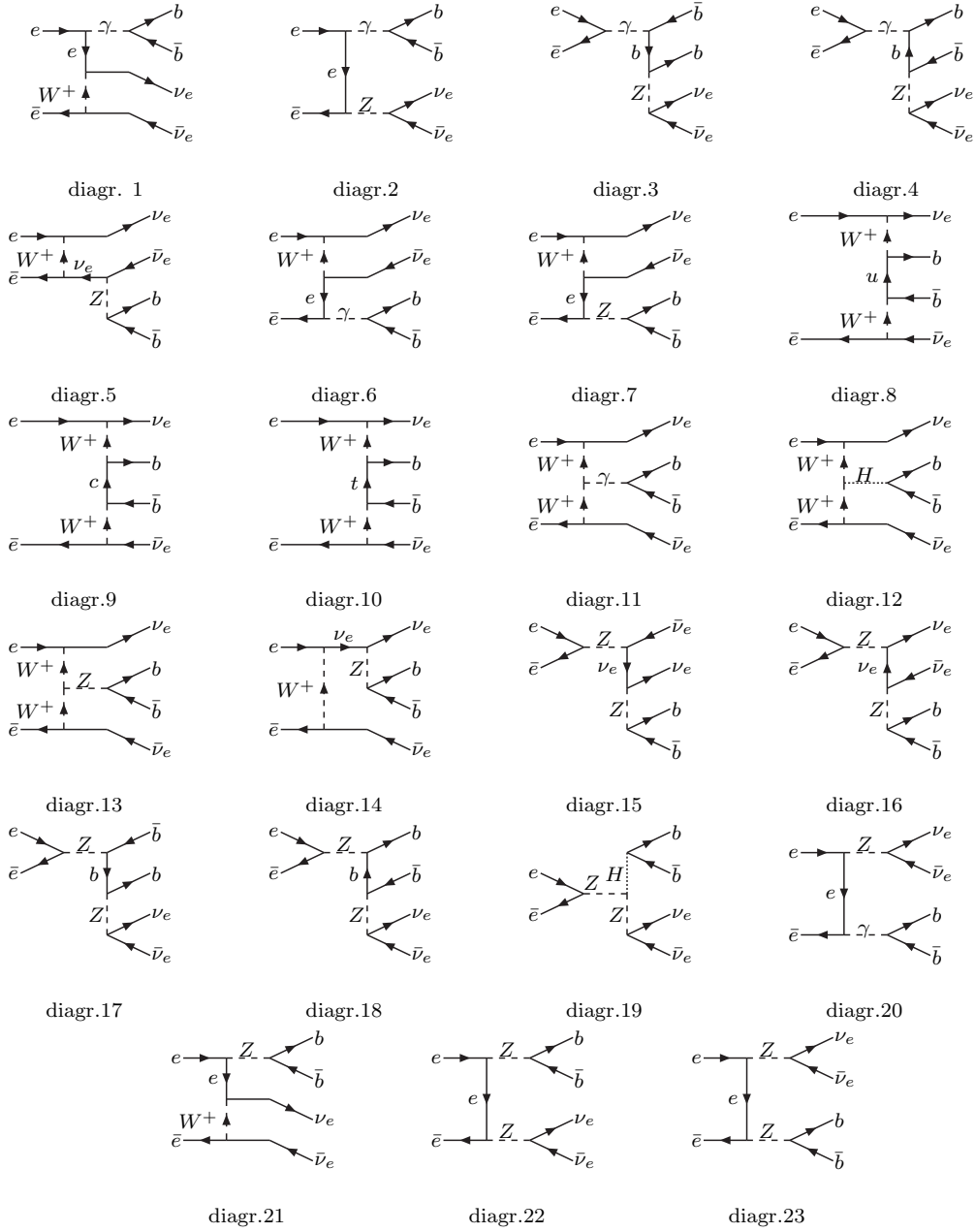

If, however, the Higgs mass is below 140 GeV, the
search becomes more challenging, since the backgrounds
to the $b\bar{b}\not\!\!\!E$ final state are much larger.
The dominant background contributions come
from $b\bar{b}\nu_e\bar{\nu}_e$, $b\bar{b}\nu_\mu\bar{\nu}_\mu$
and $b\bar{b}\nu_\tau\bar{\nu}_\tau$ final states.
Including the SM diagrams with a Higgs 
into the background, one has a total of 23, 11 and 11
diagrams, respectively. We show the diagrams for the process
$e^+e^-\rightarrow b\bar{b}\nu_e\bar{\nu_e}$ 
in Fig.~\ref{diagrams_NNBB}.
The diagrams for the processes
$e^+e^-\rightarrow b\bar{b}\nu_\mu\bar{\nu}_\mu$
and $e^+e^-\rightarrow b\bar{b}\nu_\tau\bar{\nu}_\tau$
are obtained by obvious replacements and
dropping the diagrams containing a $W$ boson.
Notice that most often the two b-jets are not coming from
a Higgs, so one can use a ($m_h$ dependent)
invariant mass cut on the b-jets. The optimal mass window
will depend on the features of the particular detector,
for our study here we shall take $|m_{bb}-m_h|<10$ GeV.
In addition, the availability of highly polarized beams
at the NLC can substantially suppress the background (which
necessarily involves the production of $W$ or $Z$ bosons),
since $W$'s only couple to left-handed leptons.
Finally, the $ZZ$ background
with one $Z$ decaying invisibly can be further
suppressed by requiring that the missing mass $M\!\!\!\!/$
in the interaction is not near $M_Z$.
The remaining $WWZ$ background is small and can
be further reduced by a lepton veto.

\begin{figure}[t!]
\centerline{\psfig{figure=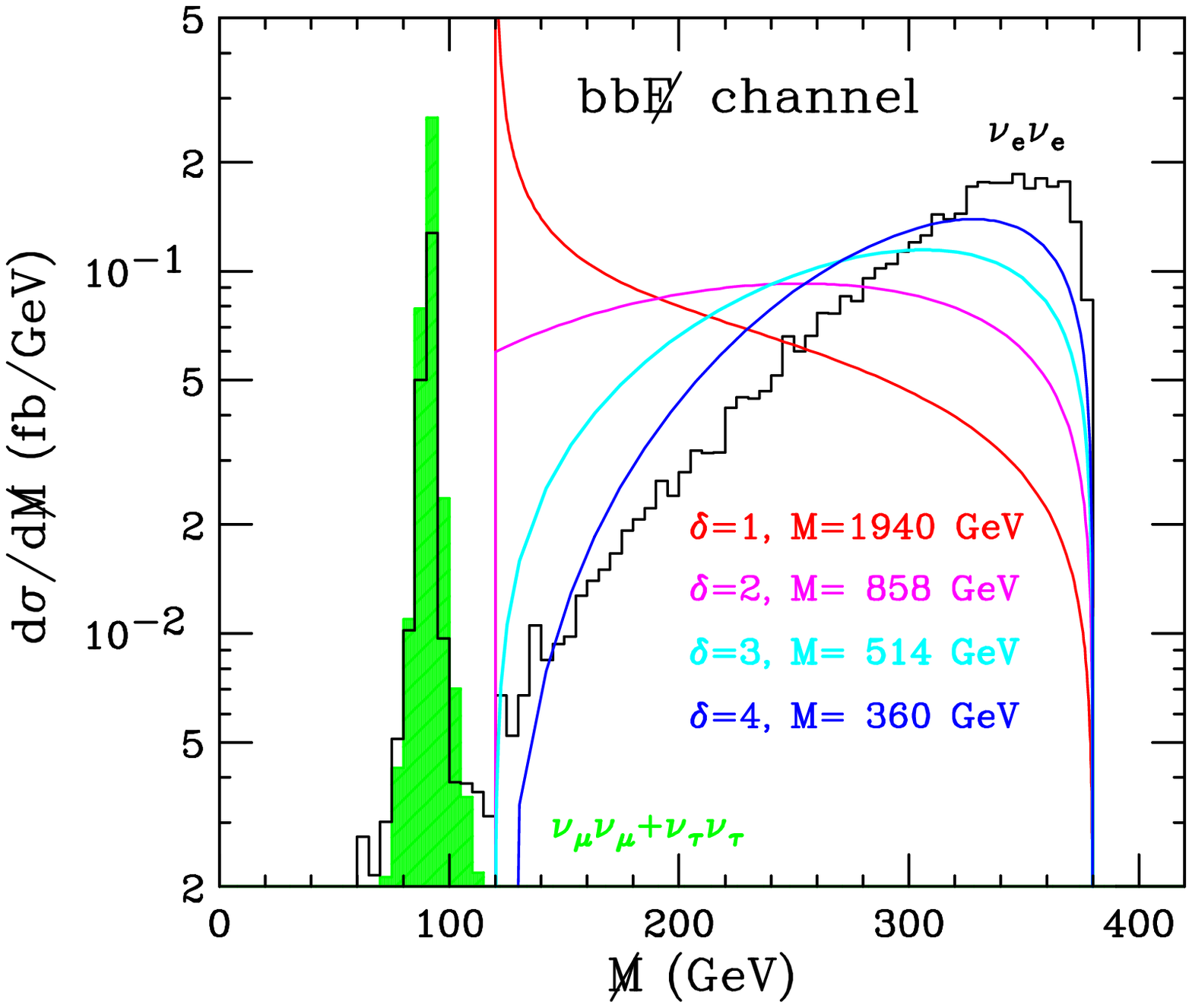,height=3.5in}}
\begin{center}
\parbox{5.5in}{
\caption[] {\small Missing mass distribution of signal
and $e^+e^-\rightarrow b\bar{b}\nu_i\bar{\nu}_i$
background at the NLC with center-of-mass energy $\sqrt{s}=500$ GeV.
The signal distributions are shown for $m_h=120$ GeV, $m_0=120$ GeV,
and several values of $\delta$ and $M$:
(a) $\delta=1$, $M=1940$ GeV, (b) $\delta=2$, $M=858$ GeV and
(c) $\delta=3$, $M=514$ GeV, (d) $\delta=4$, $M=360$ GeV.
All lines are normalized to the correct total cross-section.
We have assumed 90\% polarization and imposed the cut
$|m_{bb}-m_h|<10$ GeV. We show the $b\bar{b}\nu_e\bar{\nu}_e$
background separately from $b\bar{b}\nu_\mu\bar{\nu}_\mu
+b\bar{b}\nu_\tau\bar{\nu}_\tau$ (shaded).
\label{missing mass}}}
\end{center}
\end{figure}

We used the COMPHEP event generator \cite{COMPHEP},
to simulate the $b\bar{b}\nu\bar{\nu}$ backgrounds
(with separate runs for electron and muon/tau neutrinos)
and in Fig.~\ref{missing mass} show the resulting
missing mass $\not\!\!\!M$ distribution
at the NLC with $\sqrt{s}=500$ GeV, where
we assume a $90\%$ right-polarized electron beam.
We also present signal distributions for $m_h=120$ GeV,
$m_0=120$ GeV and several values of $\delta$ and $M$:
(a) $\delta=1$, $M=1940$ GeV,
(b) $\delta=2$, $M=858$ GeV, (b) $\delta=3$, $M=514$ GeV and
(c) $\delta=4$, $M=360$ GeV, where we have picked $M$
so that the total signal cross-section is always 20 fb.
We choose to do this since our objective here is to compare
the shape of the signal distributions to the shape of the background.

We see from Fig.~\ref{missing mass} that the shape of the
missing mass distribution for the signal depends on the 
number of extra dimensions $\delta$. In particular, for $\delta=1$
it exhibits a narrow asymmetric peak near $m_0$, for 
$\delta=2$ it is rather flat, while for larger $\delta$ it
starts to peak closer to the largest kinematically allowed
values and therefore becomes very similar to the background
distribution. (This behavior is easy to understand in terms of the
density function -- see eq.~(\ref{mass density}) from the Appendix.)
Because of the finite detector resolution, the narrow
one-sided peak for $\delta=1$ will be smeared and the flavon will
look like an ordinary heavy, invisibly decaying
neutral particle, so if a signal is seen, an interpretation 
in terms of the scenario considered here will not be
straightforward. Notice that there is very small 
background with $\not\!\!\!M<M_Z$, so by far the
easiest case to observe will be $\delta=1$ with $m_0<M_Z$,
since signal and background are very well separated.

\subsection{Hadron colliders}

In Run II of the Tevatron one can look for a $b\bar{b}\met$
final state, which is one of the standard Higgs search
channels \cite{ZHtoNuNubb}, but without mass reconstruction
may also be indicative of
several other sources of new physics:
bottom squark production \cite{sbottom},
stop production with Higgsino LSP \cite{stop},
leptoquarks \cite{leptoquark} etc.
We can adapt the existing Run II Higgs analysis of
$Zh\rightarrow b\bar{b}\met$ for our purposes,
keeping in mind that further optimization is possible, e.g.
by adjusting the $\met$ cut and the $p_T$ cuts on the b-jets.

In order to make an estimate of the signal cross-section
at the Tevatron, let us first switch the order of
integration in (\ref{hadronic xsec}):
\beq
\sigma_{tot}(E_{\rm CM})\ = \
\int_{m_0}^{E_{\rm CM}-m_h} dm_{\bf n}\,
2m_{\bf n} \rho(m_{\bf n}) 
\int_{x(\mchi)}^1 dx_1 \int_{x(\mchi)\over x_1}^1 dx_2
f^{p}(x_1) f^{\bar{p}}(x_2) 
\sigma_{\bf n}(\mchi,x_1x_2E^2_{CM})
\eeq
where $x(\mchi)\equiv (\mchi+m_h)^2/E^2_{CM}$.
Then, as long as we are interested in the total signal
cross-section only, and not in the resulting
kinematic distributions, we can use COMPHEP, for example,
to first compute the differential cross-section
at a given $\mchi$
\beq
\sigma_{\bf n}(\mchi,E_{\rm CM})\ \equiv\
{1\over\rho(\mchi)}{d\sigma_{tot}\over dm_{\bf n}^2}\ = \
\int_{x(\mchi)}^1 dx_1 \int_{{x(\mchi)\over x_1}}^1 dx_2
f^{p}(x_1) f^{\bar{p}}(x_2) 
\sigma_{\bf n}(\mchi,x_1x_2E^2_{CM})
\nonumber
\eeq
and then numerically integrate over $\mchi$.

\begin{figure}[t!]
\centerline{\psfig{figure=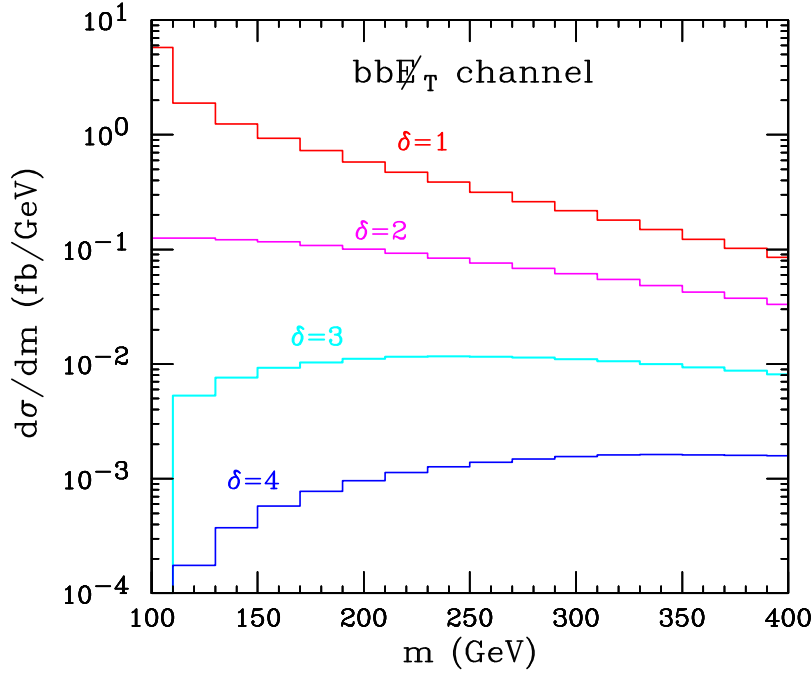,height=3.5in}}
\begin{center}
\parbox{5.5in}{
\caption[] {\small Plot of the signal cross-section 
for real flavon emission at the Tevatron for $m_h=120$ GeV,
$m_0=100$ GeV, $M=500$ GeV and several values of $\delta$.
\label{dsig_dm}}}
\end{center}
\end{figure}

In Fig.~\ref{dsig_dm} we show a plot of the differential signal
cross-section $\sigma_{\bf n}(\mchi,2\, {\rm TeV})$ before cuts. 
Here we include the signal contribution only from real flavon emission
and not from $Zh$. Notice that depending on the flavor of the
quarks participating in the hard scattering, different species of
flavons will be produced. However, we assume that they all escape 
the detector and give rise to $\met$. For the plot we have
made the simplifying assumption that the couplings $y_f$
are universal and that the mass of the zero mode $m_0$ is the same
for all species of flavons.

In Fig.~\ref{dsig_dm} we observe the typical dependence
of the mass distribution on the number of extra dimensions
$\delta$, already familiar from Fig.~\ref{missing mass}.
We see that for $\delta>2$ the distributions peak at
rather large values of $\mchi$ of around 300-400 GeV.
Notice, however, that although the actual missing
energy in those events will be very large (typically
on the order of a few hundred GeV), the distribution of the
{\em measured} $\met$ will be much softer.
The measured $\met$ in the detector is
determined by the energy of the visible products,
in our case the two b-jets, which is not much above $m_h/2$.
Nevertheless, we remind the reader that the Higgs search
in Run IIb will be sensitive to raw $b\bar{b}\met$ cross-sections
on the order of 26 fb \cite{ZHtoNuNubb}. We then see from
Fig.~\ref{dsig_dm} that flavon searches in the same channel
may yield very strong bounds on $M$, for example,
we can see from the plot that $M \lsim 1000,\, 500,\, 300,\, 200$ GeV
for $\delta=1,\, 2,\, 3,\, 4$ respectively
can be easily ruled out.

At a $pp$ collision machine like LHC, the initial
state anti-fermion must come from a sea quark, which
will suppress the production cross-section.
At the LHC it will be very hard to observe the $b\bar{b}\met$ final state,
instead one typically has to utilize the decays $h\rightarrow\gamma\gamma$
and/or $h\rightarrow ZZ$. The possibility of producing an additional
invisible flavon state in the interaction gives rise to unique clean
signatures like $\gamma\gamma\met$ (for low Higgs masses)
or $\ell\ell\ell\ell\met$ (for higher Higgs masses).

\section{Virtual Flavon exchange}
\label{virtual}

Throughout this section we shall assume that
each flavon only couples to
the corresponding pair of fermions.
Then, virtual flavon exchange at an $e^+e^-$
machine can only give back final states with electrons and positrons,
while flavon exchange at a hadron collider will yield signatures
with jets.

Virtual flavons can mediate processes with
2-fermion, 4-fermion or 6-fermion final states.
The corresponding Feynman diagrams for an $e^+e^-$ lepton
collider are shown in Figs.~\ref{diagrams2}, \ref{diagrams4}
and \ref{diagrams6}, respectively
(for a hadron collider, just replace $e\rightarrow q$).
\begin{figure}[t!]
\begin{center}
{
\unitlength=1.2 pt
\SetScale{1.2}
\SetWidth{1.0}      
\normalsize    
{} \qquad\allowbreak
\begin{picture}(95,79)(0,0)
\Text(15.0,70.0)[r]{$e$}
\ArrowLine(16.0,70.0)(37.0,60.0) 
\Text(15.0,50.0)[r]{$\bar{e}$}
\ArrowLine(37.0,60.0)(16.0,50.0) 
\Text(47.0,62.0)[b]{$\chi$}
\DashLine(37.0,60.0)(58.0,60.0){1.0}
\Text(80.0,70.0)[l]{$e$}
\ArrowLine(58.0,60.0)(79.0,70.0) 
\Text(80.0,50.0)[l]{$\bar{e}$}
\ArrowLine(79.0,50.0)(58.0,60.0) 
\Text(47,20)[b] {(a)}
\end{picture} \ 
{} \qquad\allowbreak
\begin{picture}(95,79)(0,0)
\Text(15.0,70.0)[r]{$e$}
\ArrowLine(16.0,70.0)(58.0,70.0) 
\Text(80.0,70.0)[l]{$e$}
\ArrowLine(58.0,70.0)(79.0,70.0) 
\Text(57.0,60.0)[r]{$\chi$}
\DashLine(58.0,70.0)(58.0,50.0){1.0}
\Text(15.0,50.0)[r]{$\bar{e}$}
\ArrowLine(58.0,50.0)(16.0,50.0) 
\Text(80.0,50.0)[l]{$\bar{e}$}
\ArrowLine(79.0,50.0)(58.0,50.0) 
\Text(47,20)[b] {(b)}
\end{picture} \ 
}
\parbox{5.5in}{
\caption[] {\small The diagrams for the process
$e^+e^-\rightarrow \chi^\ast \rightarrow e^+ e^-$.
\label{diagrams2}}}
\end{center}
\end{figure}
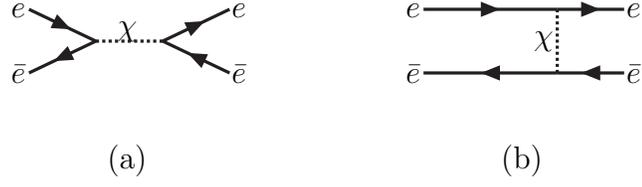
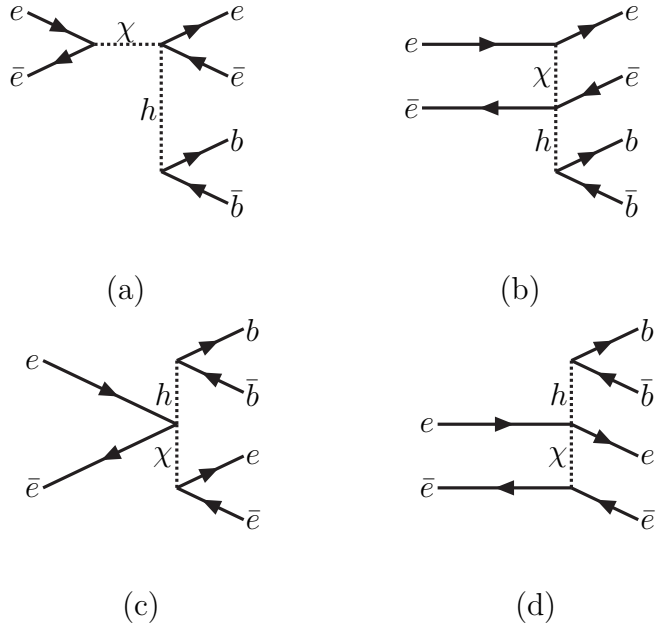
\begin{figure}[t!]
\begin{center}
{
\unitlength=1.2 pt
\SetScale{1.2}
\SetWidth{1.0}      
\normalsize    
{} \qquad\allowbreak
\begin{picture}(95,99)(0,0)
\Text(15.0,90.0)[r]{$e$}
\ArrowLine(16.0,90.0)(37.0,80.0) 
\Text(15.0,70.0)[r]{$\bar{e}$}
\ArrowLine(37.0,80.0)(16.0,70.0) 
\Text(47.0,83.0)[b]{$\chi$}
\DashLine(37.0,80.0)(58.0,80.0){1.0}
\Text(80.0,90.0)[l]{$e$}
\ArrowLine(58.0,80.0)(79.0,90.0) 
\Text(80.0,70.0)[l]{$\bar{e}$}
\ArrowLine(79.0,70.0)(58.0,80.0) 
\Text(57.0,60.0)[r]{$h$}
\DashLine(58.0,80.0)(58.0,40.0){1.0}
\Text(80.0,50.0)[l]{$b$}
\ArrowLine(58.0,40.0)(79.0,50.0) 
\Text(80.0,30.0)[l]{$\bar{b}$}
\ArrowLine(79.0,30.0)(58.0,40.0) 
\Text(47,0)[b] {(a)}
\end{picture} \ 
{} \qquad\allowbreak
\begin{picture}(95,99)(0,0)
\Text(15.0,80.0)[r]{$e$}
\ArrowLine(16.0,80.0)(58.0,80.0) 
\Text(80.0,90.0)[l]{$e$}
\ArrowLine(58.0,80.0)(79.0,90.0) 
\Text(57.0,70.0)[r]{$\chi$}
\DashLine(58.0,80.0)(58.0,60.0){1.0}
\Text(15.0,60.0)[r]{$\bar{e}$}
\ArrowLine(58.0,60.0)(16.0,60.0) 
\Text(80.0,70.0)[l]{$\bar{e}$}
\ArrowLine(79.0,70.0)(58.0,60.0) 
\Text(57.0,50.0)[r]{$h$}
\DashLine(58.0,60.0)(58.0,40.0){1.0}
\Text(80.0,50.0)[l]{$b$}
\ArrowLine(58.0,40.0)(79.0,50.0) 
\Text(80.0,30.0)[l]{$\bar{b}$}
\ArrowLine(79.0,30.0)(58.0,40.0) 
\Text(47,0)[b] {(b)}
\end{picture} \ 
{} \qquad\allowbreak
\\
\begin{picture}(95,99)(0,0)
\Text(15.0,80.0)[r]{$e$}
\ArrowLine(16.0,80.0)(58.0,60.0) 
\Text(15.0,40.0)[r]{$\bar{e}$}
\ArrowLine(58.0,60.0)(16.0,40.0) 
\Text(57.0,70.0)[r]{$h$}
\DashLine(58.0,60.0)(58.0,80.0){1.0}
\Text(80.0,90.0)[l]{$b$}
\ArrowLine(58.0,80.0)(79.0,90.0) 
\Text(80.0,70.0)[l]{$\bar{b}$}
\ArrowLine(79.0,70.0)(58.0,80.0) 
\Text(57.0,50.0)[r]{$\chi$}
\DashLine(58.0,60.0)(58.0,40.0){1.0}
\Text(80.0,50.0)[l]{$e$}
\ArrowLine(58.0,40.0)(79.0,50.0) 
\Text(80.0,30.0)[l]{$\bar{e}$}
\ArrowLine(79.0,30.0)(58.0,40.0) 
\Text(47,0)[b] {(c)}
\end{picture} \ 
{} \qquad\allowbreak
\begin{picture}(95,99)(0,0)
\Text(15.0,60.0)[r]{$e$}
\ArrowLine(16.0,60.0)(58.0,60.0) 
\Text(57.0,70.0)[r]{$h$}
\DashLine(58.0,60.0)(58.0,80.0){1.0}
\Text(80.0,90.0)[l]{$b$}
\ArrowLine(58.0,80.0)(79.0,90.0) 
\Text(80.0,70.0)[l]{$\bar{b}$}
\ArrowLine(79.0,70.0)(58.0,80.0) 
\Text(80.0,50.0)[l]{$e$}
\ArrowLine(58.0,60.0)(79.0,50.0) 
\Text(57.0,50.0)[r]{$\chi$}
\DashLine(58.0,60.0)(58.0,40.0){1.0}
\Text(15.0,40.0)[r]{$\bar{e}$}
\ArrowLine(58.0,40.0)(16.0,40.0) 
\Text(80.0,30.0)[l]{$\bar{e}$}
\ArrowLine(79.0,30.0)(58.0,40.0) 
\Text(47,0)[b] {(d)}
\end{picture} \ 
}
\parbox{5.5in}{
\caption[] {\small The diagrams for the process
$e^+e^-\rightarrow \chi^\ast \rightarrow e^+e^-b\bar{b}$.
\label{diagrams4}}}
\end{center}
\end{figure}
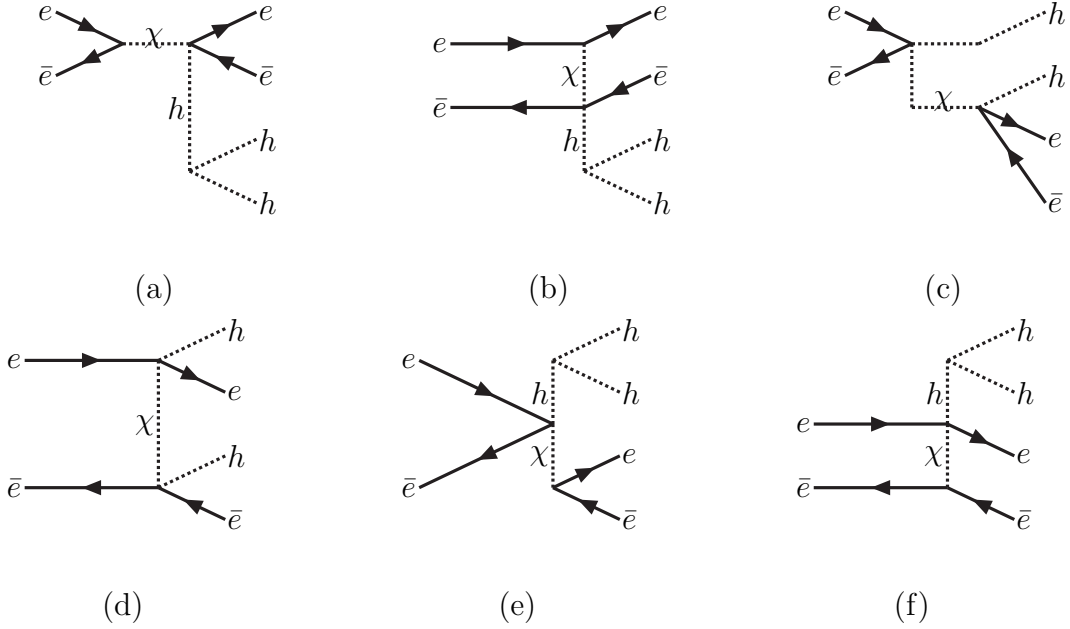
\begin{figure}[t!]
\begin{center}
{
\unitlength=1.2 pt
\SetScale{1.2}
\SetWidth{1.0}      
\normalsize    
{} \qquad\allowbreak
\begin{picture}(95,99)(0,0)
\Text(15.0,90.0)[r]{$e$}
\ArrowLine(16.0,90.0)(37.0,80.0) 
\Text(15.0,70.0)[r]{$\bar{e}$}
\ArrowLine(37.0,80.0)(16.0,70.0) 
\Text(47.0,81.0)[b]{$\chi$}
\DashLine(37.0,80.0)(58.0,80.0){1.0}
\Text(80.0,90.0)[l]{$e$}
\ArrowLine(58.0,80.0)(79.0,90.0) 
\Text(80.0,70.0)[l]{$\bar{e}$}
\ArrowLine(79.0,70.0)(58.0,80.0) 
\Text(57.0,60.0)[r]{$h$}
\DashLine(58.0,80.0)(58.0,40.0){1.0}
\Text(80.0,50.0)[l]{$h$}
\DashLine(58.0,40.0)(79.0,50.0){1.0}
\Text(80.0,30.0)[l]{$h$}
\DashLine(58.0,40.0)(79.0,30.0){1.0}
\Text(47,0)[b] {(a)}
\end{picture} \ 
{} \qquad\allowbreak
\begin{picture}(95,99)(0,0)
\Text(15.0,80.0)[r]{$e$}
\ArrowLine(16.0,80.0)(58.0,80.0) 
\Text(80.0,90.0)[l]{$e$}
\ArrowLine(58.0,80.0)(79.0,90.0) 
\Text(57.0,70.0)[r]{$\chi$}
\DashLine(58.0,80.0)(58.0,60.0){1.0}
\Text(15.0,60.0)[r]{$\bar{e}$}
\ArrowLine(58.0,60.0)(16.0,60.0) 
\Text(80.0,70.0)[l]{$\bar{e}$}
\ArrowLine(79.0,70.0)(58.0,60.0) 
\Text(57.0,50.0)[r]{$h$}
\DashLine(58.0,60.0)(58.0,40.0){1.0}
\Text(80.0,50.0)[l]{$h$}
\DashLine(58.0,40.0)(79.0,50.0){1.0}
\Text(80.0,30.0)[l]{$h$}
\DashLine(58.0,40.0)(79.0,30.0){1.0}
\Text(47,0)[b] {(b)}
\end{picture} \ 
{} \qquad\allowbreak
\begin{picture}(95,99)(0,0)
\Text(15.0,90.0)[r]{$e$}
\ArrowLine(16.0,90.0)(37.0,80.0) 
\Text(15.0,70.0)[r]{$\bar{e}$}
\ArrowLine(37.0,80.0)(16.0,70.0) 
\DashLine(37.0,80.0)(58.0,80.0){1.0}
\Text(80.0,90.0)[l]{$h$}
\DashLine(58.0,80.0)(79.0,90.0){1.0}
\DashLine(37.0,80.0)(37.0,60.0){1.0}
\Text(47.0,61.0)[b]{$\chi$}
\DashLine(37.0,60.0)(58.0,60.0){1.0}
\Text(80.0,70.0)[l]{$h$}
\DashLine(58.0,60.0)(79.0,70.0){1.0}
\Text(80.0,50.0)[l]{$e$}
\ArrowLine(58.0,60.0)(79.0,50.0) 
\Text(80.0,30.0)[l]{$\bar{e}$}
\ArrowLine(79.0,30.0)(58.0,60.0) 
\Text(47,0)[b] {(c)}
\end{picture} \ 
{} \qquad\allowbreak
\begin{picture}(95,99)(0,0)
\Text(15.0,80.0)[r]{$e$}
\ArrowLine(16.0,80.0)(58.0,80.0) 
\Text(80.0,90.0)[l]{$h$}
\DashLine(58.0,80.0)(79.0,90.0){1.0}
\Text(80.0,70.0)[l]{$e$}
\ArrowLine(58.0,80.0)(79.0,70.0) 
\Text(57.0,60.0)[r]{$\chi$}
\DashLine(58.0,80.0)(58.0,40.0){1.0}
\Text(15.0,40.0)[r]{$\bar{e}$}
\ArrowLine(58.0,40.0)(16.0,40.0) 
\Text(80.0,50.0)[l]{$h$}
\DashLine(58.0,40.0)(79.0,50.0){1.0}
\Text(80.0,30.0)[l]{$\bar{e}$}
\ArrowLine(79.0,30.0)(58.0,40.0) 
\Text(47,0)[b] {(d)}
\end{picture} \ 
{} \qquad\allowbreak
\begin{picture}(95,99)(0,0)
\Text(15.0,80.0)[r]{$e$}
\ArrowLine(16.0,80.0)(58.0,60.0) 
\Text(15.0,40.0)[r]{$\bar{e}$}
\ArrowLine(58.0,60.0)(16.0,40.0) 
\Text(57.0,70.0)[r]{$h$}
\DashLine(58.0,60.0)(58.0,80.0){1.0}
\Text(80.0,90.0)[l]{$h$}
\DashLine(58.0,80.0)(79.0,90.0){1.0}
\Text(80.0,70.0)[l]{$h$}
\DashLine(58.0,80.0)(79.0,70.0){1.0}
\Text(57.0,50.0)[r]{$\chi$}
\DashLine(58.0,60.0)(58.0,40.0){1.0}
\Text(80.0,50.0)[l]{$e$}
\ArrowLine(58.0,40.0)(79.0,50.0) 
\Text(80.0,30.0)[l]{$\bar{e}$}
\ArrowLine(79.0,30.0)(58.0,40.0) 
\Text(47,0)[b] {(e)}
\end{picture} \ 
{} \qquad\allowbreak
\begin{picture}(95,99)(0,0)
\Text(15.0,60.0)[r]{$e$}
\ArrowLine(16.0,60.0)(58.0,60.0) 
\Text(57.0,70.0)[r]{$h$}
\DashLine(58.0,60.0)(58.0,80.0){1.0}
\Text(80.0,90.0)[l]{$h$}
\DashLine(58.0,80.0)(79.0,90.0){1.0}
\Text(80.0,70.0)[l]{$h$}
\DashLine(58.0,80.0)(79.0,70.0){1.0}
\Text(80.0,50.0)[l]{$e$}
\ArrowLine(58.0,60.0)(79.0,50.0) 
\Text(57.0,50.0)[r]{$\chi$}
\DashLine(58.0,60.0)(58.0,40.0){1.0}
\Text(15.0,40.0)[r]{$\bar{e}$}
\ArrowLine(58.0,40.0)(16.0,40.0) 
\Text(80.0,30.0)[l]{$\bar{e}$}
\ArrowLine(79.0,30.0)(58.0,40.0) 
\Text(47,0)[b] {(f)}
\end{picture} \ 
}
\parbox{5.5in}{
\caption[] {\small The diagrams for the process
$e^+e^-\rightarrow \chi^\ast h\rightarrow e^+e^-hh$.
Each Higgs further decays to a $b\bar{b}$
or $\tau^+\tau^-$ pair.
\label{diagrams6}}}
\end{center}
\end{figure}

For $\delta>2$, the virtual $\chi$ exchange can be represented by a local
effective dimension 8 operator (\ref{eff operator})
as shown in the Appendix.
In Table~\ref{virtual exchanges} we list the signal cross-sections
(in fb) from this effective operator at different colliders for
$\Lambda_T=1$ TeV and $m_h=120$ GeV. 
\begin{table}[t!]
\centering
\renewcommand{\arraystretch}{1.2}
\begin{tabular}{||c||c|c||c|c||}
\hline\hline
Final
  & LEP
    & NLC
      & FNAL
        & LHC     \\  \cline{2-5}
state 
  & 200 GeV
    & 500 GeV
      & 2 TeV
        & 14 TeV  \\ \hline\hline	
$  ff$
  & 0.95
    & $5.91$
      & 12.9
        & 373     \\ \hline
$  ffh$
  & $9\times10^{-5}$
    & $6\times10^{-2}$
      & $2.8\times10^{-2}$
        & 60      \\ \hline
$  ffhh$
  & ---
    & $6.5\times10^{-6}$
      & $2.9\times10^{-4}$
        & 2.7     \\ \hline\hline
\end{tabular}
\parbox{5.5in}{
\caption{ Signal cross-sections in fb for the
2-fermion, 4-fermion ($f\bar{f}b\bar{b}$)
and 6-fermion ($f\bar{f}b\bar{b}b\bar{b}$) final states,
at various colliders (with their center of mass energy shown)
and for $\Lambda_T=1$ TeV and $m_h=120$ GeV.
\label{virtual exchanges}}}
\end{table}
In the case of a real Higgs production, we only consider
decays to $b\bar{b}$ final states, whose branching fraction
was computed with the HDECAY program \cite{hdecay}. 
If $\Lambda_c=M$ and $y_f=1$, then $\Lambda_T=1$ TeV
corresponds to $M=400,237,145$ GeV for $\delta=3,\,4,\,5$,
respectively. These cross-sections can be readily rescaled
for different values of $\Lambda_T$, recall that
$\sigma\sim\Lambda_T^{-8}$.

We see from Table~\ref{virtual exchanges}
that the signal cross-sections are rather small,
and just like in the case of virtual graviton exchanges,
one will have to look for interference effects instead,
since those are only proportional to $\Lambda_T^{-4}$.

For $\delta \leq 2$, the effective interaction still depends on
the momentum flow along the $\chi$ propagator. In the case of $\delta=2$,
there is only a mild logarithmic dependence on the momentum and
the cutoff. The case of $\delta=1$ is more interesting. The integral
over KK states is convergent, therefore independent of the cutoff.
For $p^2>m_0^2$ ($p$ is the momentum of the $\chi$ field), the
effective interaction is imaginary, corresponding to real
$\chi$ production. Whether it gives rise to the same final state particles
or not, depends on the $\chi$ lifetime. Real $\chi$ production does
not interfere with other virtual (off-shell) processes.
The singularity at $p^2=m_0^2$ will be smoothed out
by the decay width of the $\chi$ fields and the detector resolution.
Nevertheless, the peak at $p^2=m_0^2$ provides an extra handle of the
$\delta=1$ case.

In the rest of this Section we shall discuss lepton and
hadron colliders separately.

\subsection{Lepton colliders}

\begin{figure}[t!]
\centerline{\psfig{figure=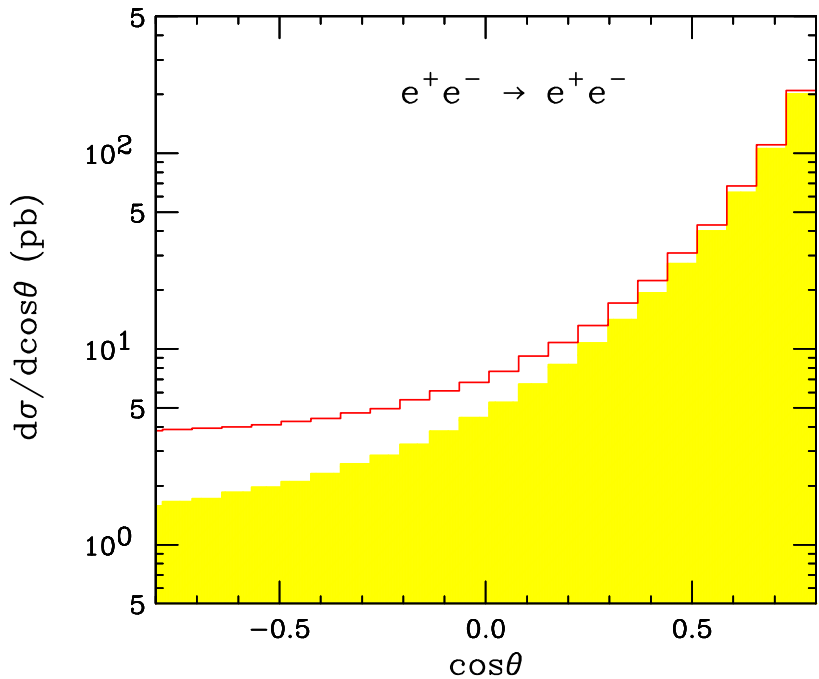,height=3.5in}}
\begin{center}
\parbox{5.5in}{
\caption[] {\small Angular distribution for both signal and
background at LEP with $\sqrt{s}=200$ GeV in Bhabha scattering,
for $\Lambda_T=750$ GeV.
\label{ang_LEP}}}
\end{center}
\end{figure}

The virtual scalar exchanges in Fig.~\ref{diagrams2} will modify
the angular distribution of the leptons in Bhabha scattering
and may be detected at LEP. In Fig.~\ref{ang_LEP} we
show the differential cross-section $d\sigma/d\cos\theta$
(in fb) at LEP for $\Lambda_T=750$ GeV,
where $\theta$ is the scattering angle of the
electron with respect to the direction of the electron beam.
The shaded histogram is just the background contribution, while
the solid histogram includes both signal and background.
There is a visible excess of backscattered electrons,
which is typical of the (isotropic) scalar exchange.
In order to extract the LEP limits on $\Lambda_T$ from
Bhabha scattering, we can apply the latest bounds on
the relevant compositeness scale $\Lambda_{\rm LR}$
\cite{compositeness}. We have to equate the
coefficient of the contact term
\beq
{g^2\over2 \Lambda^2_{\rm LR}}
(\bar{e}_L\gamma^\mu e_L)(\bar{e}_R\gamma_\mu e_R)
\sim{g^2\over \Lambda^2_{\rm LR}}(\bar{e}_L e_R)(\bar{e}_R e_L)
\eeq
to the coefficient of the effective operator
(\ref{eff operator}). The typical bounds on $\Lambda_{\rm LR}$
of 4.5 TeV for $g^2\sim 4\pi$ then translate into
\beq
\Lambda_T\gsim \left({v\Lambda_{\rm LR}
\over\sqrt{4\pi}}\right)^{1/2}
\sim 525\ {\rm GeV}.
\label{limit from compositeness}
\eeq
We should point out that in the case of $\delta=1$ and $m_0$
not near $M_Z$, one may use in addition the peak in the invariant
mass of the leptons.

\begin{figure}[t!]
\centerline{\psfig{figure=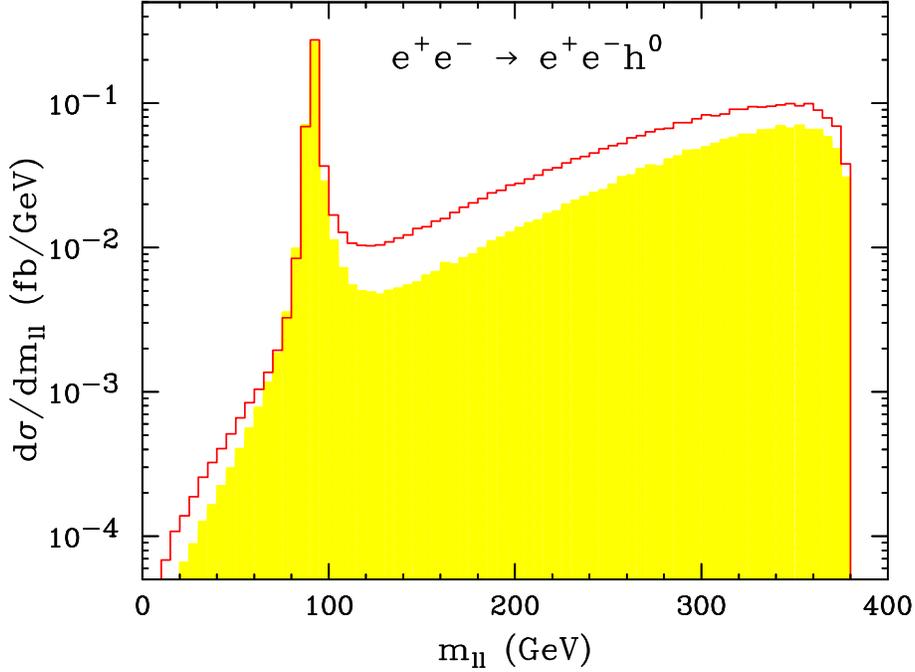,height=3.5in}}
\begin{center}
\parbox{5.5in}{
\caption[] {\small Invariant mass distribution of the $e^+e^-$
pair in $e^+e^-h^0$ events for both signal (solid) and
background (shaded) at the NLC with $\sqrt{s}=500$ GeV,
and for $\Lambda_T=600$ GeV and $m_h=120$ GeV.
\label{virt_llh_m}}}
\end{center}
\end{figure}

\begin{figure}[t!]
\centerline{\psfig{figure=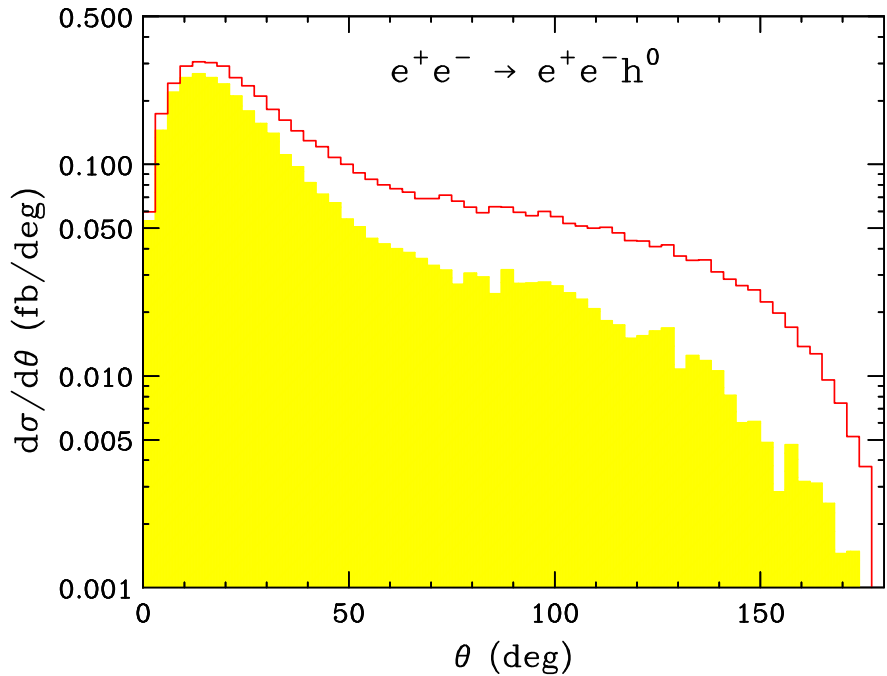,height=3.5in}}
\begin{center}
\parbox{5.5in}{
\caption[] {\small The same as Fig~\ref{virt_llh_m}, but for
the angular distribution $d\sigma/d\theta$, where
$\theta$ is the scattering angle of the electron
with respect to the electron beam.
\label{virt_llh_an}}}
\end{center}
\end{figure}

We next discuss the 4-fermion channel, which
is similar to the existing $Zh\rightarrow e^+e^-b\bar{b}$ searches
\cite{LEP Higgs searches}. Here the invariant mass $m_{e^+e^-}$
of the $e^+e^-$ pair is not constrained to be close to $M_Z$,
so one might think that $m_{e^+e^-}$ will be a useful variable
to cut on. In Fig.~\ref{virt_llh_m} we show the $m_{e^+e^-}$
distribution in $e^+e^-h^0$ events
for signal plus background (solid) or just the background (shaded)
at an NLC with $\sqrt{s}=500$ GeV,
and for $\Lambda_T=600$ GeV and $m_h=120$ GeV.
We see that $m_{e^+e^-}$ does not allow good discrimination
between signal and background, since the two distributions
are very similar. Above the $Z$-peak, the background is
dominated by $Z$-fusion, just like $W$-fusion was the
main background for the $b\bar{b}\met$ channel in
Section~\ref{direct-lepton colliders}. Unlike $W$-fusion,
here we have the additional advantage that we can measure the
angular distribution of the individual leptons. This might be
useful because the leptons in $Z$-fusion tend to be forward,
while the leptons from the signal are isotropic.
This is illustrated in Fig.~\ref{virt_llh_an}, where
we show the angular distribution of the electron
with respect to the electron beam for the same
parameters as in Fig.~\ref{virt_llh_m}. We see that
$\theta$ can be a decent discriminator between
signal and background. 

And finally, given the current limits on $m_h$, the 6-fermion final
state is not observable at LEP. From Table~\ref{virtual exchanges}
we estimate that we can record 5 signal events at the NLC with
$\sqrt{s}=500$ GeV and total integrated luminosity 50 ${\rm fb}^{-1}$,
only if $\Lambda_T < 300$ GeV. Such low values of $\Lambda_T$
are already ruled out by LEP
(see eq.~(\ref{limit from compositeness}) above).

\subsection{Hadron colliders}

At hadron colliders we can similarly have 2-jet,
4-jet and 6-jet final states, by replacing
$e\rightarrow q$ in Figs.~\ref{diagrams2},
\ref{diagrams4} and \ref{diagrams6}.

The dijet signal should be studied analogously to
the graviton exchange \cite{dijets}, the difference
now being that the angular distributions should be
characteristic of spin 0 rather than spin 2 virtual exchange.

The observation of the 4-jet final state will be rather difficult.
There will be large irreducible backgrounds from $t\bar{t}$,
$Z+2j$, $WZ$ and $ZZ$, besides, the signal looks just like
$gg\rightarrow h$ with extra radiated jets. 
In any case, the presence of extra jets is not very useful
at a hadron machine. Even for larger Higgs masses,
where the Higgs decays to gauge bosons, our signal is not
distinctive enough by itself and will be simply added to
the samples for the regular Higgs search.

The 6-jet final state is cleaner, but the cross-section is
too small to be relevant for the Tevatron. However, 
Higgs pair-production processes have small Standard
Model backgrounds. Therefore identification of both Higgs
particles in the event can be an important test for the
presence of some new physics (see, e.g. \cite{rizzo}). 
The main Higgs pair-production process at the LHC is
gluon fusion \cite{Hpair}. In the four b-jet channel,
very high $p_T$ cuts on the jets are necessary in order to
suppress the QCD backgrounds, and then the acceptance is only
1-2\%, assuming rather optimistic b-tagging efficiencies
\cite{4b}. Therefore, for low Higgs masses it seems rather
difficult to see a signal, and almost impossible to distinguish
it from $hb\bar{b}$ at large $\tan\beta$. For heavy Higgs masses,
where the cross-section is smaller than the one shown in
Table~\ref{virtual exchanges}, we have 4 gauge bosons $+$ 2 jet
signatures. Then, for the purposes of both triggering and
reducing the background, one would have to use the
leptonic decays of a sufficient number of $W$'s and $Z$'s,
thus paying a significant price in branching ratios.

\section{Conclusions}
\label{conclusions}

The possibility that new large compact dimensions may be probed in the
near-term future at collider experiments is very exciting.
Finding KK excitations will provide a strong evidence for
the existence of extra dimensions.
On general grounds, one expects that in addition to
gravitons, there are other particles feeling
the extra dimensions. If they are some SM fields,
then looking for their KK excitations will provide the strongest
probe of the extra dimensions. The experimental constraints on the
masses of the KK modes of the SM fields
require the corresponding extra dimensions
to have sizes smaller than $\sim {\rm TeV}^{-1}$.
If we assume that all SM gauge fields
are localized on our brane, any bulk field must
be neutral under the SM gauge group.
Hence bulk fields will only couple to some neutral
combination of the SM fields.
In the case of a bulk fermion, it may
couple to $\bar{\ell}_L \tilde{H}$ and be identified as 
right-handed neutrino. The smallness of the neutrino masses can 
then be explained by the large size of the extra 
dimensions~\cite{ADDM,DDGn}.
It may also give rise to invisible Higgs
decays~\cite{MW}. If the bulk field is a scalar, it can
couple to neutral combinations of the SM fields such as
$\bar{f}_L f_R H$, $F^{\mu\nu} F_{\mu\nu}$, or higher dimensional
ones $QQQL$ and so on. The study of this paper can apply to any
scalar bulk fields which couple to $\bar{f}_L f_R H$, independent
of whether there is a flavor theory. The scalars which couple
to $F^{\mu\nu} F_{\mu\nu}$ are similar to the dilatons
and will be left for future studies. The scalars which couple
to higher dimensional SM operators like $QQQL$,
have suppressed interactions and are therefore
much more difficult to produce in collider experiments.

The new signals studied in this paper
are not limited to the ideas of large extra dimensions.
Ignoring for the moment the difficulties in
flavor model building, the results are also applicable
in the usual four dimensional flavor theories if the flavons are
light. In spite of the differences in the
invariant mass distribution of the flavon
(or its decay products), it seems that distinguishing
between a four dimensional and a higher dimensional
flavon scenario is intrinsically quite difficult
and may have to wait for even more advanced experimental
facilities.

To summarize, we have studied many new signals coming from
flavor physics within the framework of large extra
dimensions and TeV scale gravity. These new signatures
are particularly interesting because they are
associated with the Higgs boson, which is the only missing piece
of the Standard Model and hence the most hunted particle.
We find that the production of Higgs + real flavon could probe
the string scale just as well as the KK graviton production
at the current and future colliders, if the flavons and the graviton
live in the same number of extra dimesions. However, if the flavons only
live in a subspace of the extra dimensions accesible to the graviton,
associated flavon-Higgs production can provide stronger
constraints than graviton production. In addition, there is a range
of the parameter space in which the flavons can have macroscopic
decay lengths, which will give rise to the spectacular signals of
displaced vertices. The virtual flavon exchange generates higher
dimensional operators which contain Higgs fields. However, they
do not provide as good probes as the real flavon emission, because
the backgrounds are very large for final states with 0 or 1 Higgs
bosons, while the clean signals of 2 Higgs final states are too
small to be observed.
A possible exception is the case in which the flavons only live in one
extra dimension. The fermion pair invariant mass
peaks at the mass of the flavon zero mode, providing a very
distinctive signal. This is a novel feature of the flavons in
one extra dimension, since one extra dimension for the graviton
in the TeV scale gravity scenario is ruled out. Note that in our
study we have made the conservative assumption that each flavon only
couples to one corresponding fermion pair. There could be more promising
signals if one relaxes this assumption. For instance, if a flavon couples
to both quarks and leptons, then at a hadron collider we can have 
$b\bar{b}$ (from $h$) + leptons final states, which is an
exceptionally clean signal.

\vspace*{1.5cm}

{\it Acknowledgements:} We would like to thank N.~Arkani-Hamed,
J.~Bagger, S.~Dimopoulos, L.~Hall, G.~Landsberg
and J.~Lykken for stimulating discussions.
Fermilab is operated by the URA under DOE contract DE-AC02-76CH03000.
\newpage

\appendix

\section{Formalism and conventions}
\label{formulas}

Expanding $H$ and $\chi$ around their vevs, $H=(0,\,(v+h)/\sqrt{2})$
(setting the would-be-Goldstone fields to zero)
and $\chi_{\bf n}=\VEV{\chi_{\bf n}}+ \chi'_{\bf n}$, the interactions
(\ref{KKint}) contain
\beq
{\cal H}_{\rm int} \supset
m_f {\bar f}_L f_R + {\lambda_f\over \sqrt{2}} {\bar f}_L f_R h
+ \sum_{\bf n} {y_f v\over \sqrt{2}M_F} {\bar f}_L f_R \chi'_{\bf n}
+ \sum_{\bf n} {y_f \over \sqrt{2}M_F} {\bar f}_L f_R h \chi'_{\bf n}
+ {\rm h.c.},
\eeq
where $v=246$ GeV is the Higgs vev, $m_f$ and $\lambda_f$ are
the mass and Yukawa coupling for the corresponding fermion,
$h$ is the real Higgs field and $\chi'_{\bf n}$ are complex
KK states. For simplicity, we assume that the masses of the real 
and the imaginary parts of $\chi'$ are the same. The mass for the
KK state $\chi_{\bf n}$ is given by 
$m_{\bf n}^2 =m_0^2 + 4\pi^2 |{\bf n}|^2/L^2$, where $m_0$ 
is the mass of the zero mode. Also, for notational simplicity,
we will omit the prime on $\chi'$ for the most part of the paper:
$\chi$ is understood as the vev shifted field when there is
no confusion.

We follow the approach of Han, Lykken, and Zhang~\cite{HLZ}, 
changing slightly their notation: $R\rightarrow L$ for the
{\em length} of the extra dimensions and $n\rightarrow\delta$
for the {\em number} of extra dimensions accessible to $\chi$'s.
Because there is a mass $m_0$ for the zero mode, the number of
KK states of $\chi$ in the mass interval $dm_{\bf n}^2$ is now given by
\beq
dN\approx \rho(m_{\bf n}) dm_{\bf n}^2,
\eeq
with
\beq
\rho(m_{\bf_n})= {L^\delta (m_{\bf n}^2-m_0^2)^{(\delta-2)/ 2}\over
(4\pi)^{\delta/2} \Gamma(\delta/2)}.
\label{mass density}
\eeq
The effective interaction due to the virtual $\chi$ KK state exchange
can be obtained by summing over the KK state propagator, in a way
similar to the virtual KK graviton exchange~\cite{GRW,HLZ,Hew}.
For $\delta\geq 2$
the virtual exchange is dominated by the high KK states, it depends
sensitively on the cutoff scale $\Lambda_c$, but not on $m_0$,
as long as $m_0$ is much smaller than $\Lambda_c$ (and $\sqrt{s},\,
\sqrt{|t|}$ for $\delta=2$). The sum over the KK mode propagators
gives
\bear
D(s)&=&
\int_{m_0^2}^{\Lambda_c^2} \
dm^2_{\bf n}\ \rho(m_{\bf n})\
{i\over s-m^2_{\bf n}+i\varepsilon}  \\[2mm]
&\approx&
\left\{
\ba{l}
{L^2\over 4\pi}\ \biggl[ \pi - i
\ln\biggl({\Lambda^2_c-s\over s-m_0^2} \biggr)\biggr],  
\quad \delta=2; \\[2mm]
{2 L^\delta (s-m_0^2)^{(\delta-2)/2} \over 
(\delta-2)\Gamma(\delta/2) (4\pi)^{\delta/2}}
\Biggl[ \pi -i
\biggl({\Lambda_c^2-m_0^2 \over s-m_0^2}\biggr)^{(\delta-2)/2}\Biggr], 
\quad \delta>2.
\ea
\right. 
\nonumber
\eear
The real part represents the real KK mode production. It 
contributes to the same processes as the virtual exchange
only if the KK mode of $\chi$ has prompt decay. It is smaller
than the virtual contribution if $\Lambda_c \gg \sqrt{s},\, m_0$
for $\delta \geq 2$.
The $t$-channel exchange can be obtained similarly by replacing
$s$ by $t$ (remembering that $t<0$).

The virtual exchange for $\delta >2$ is independent of $s$ and
therefore can be written as an effective dimension 8 operator,
assuming $\Lambda_c \gg m_0$ and using $M_F^2= M^{2+\delta} L^{\delta}$,
\bear
\label{EffOp}
{\cal H}_{\rm eff} &=& -{|y_f|^2\over (\delta-2)\Gamma(\delta/2)
(4\pi)^{\delta/2}}
{\Lambda_c^{\delta-2} \over M^{2+\delta}} 
|\bar{f}_L f_R h|^2  \nonumber \\
&\equiv & -
{1\over \Lambda_T^4} |\bar{f}_L f_R h|^2,
\label{eff operator}
\eear
Experiments on virtual exchange  only probe the coefficient of 
this effective operator,
$\Lambda_T^4$, which is a combination of the gravitation scale $M$
and the cutoff $\Lambda_c$.
Because the cutoff is not known, the experimental probes of
virtual exchange processes should not be directly
compared to the real $\chi$ production.

For $\delta=2$, the virtual $\chi$ exchange still depends logarithmically
on the momentum flow through $\chi$. In analogy to (\ref{EffOp}),
we will write the effective interaction as
\bear
-{|y_f|^2\over 8\pi M^4} \ln\biggl| {\Lambda_c^2-p^2\over p^2-m_0^2}
\biggr| |\bar{f}_L f_R h|^2,
\eear
where $p^2$ is the momentum flow of the virtual $\chi$.

For $\delta=1$, the integral over KK propagators is convergent,
\beq
D(s)= {L\over 2} {1\over \sqrt{s-m_0^2}} + i {L\over \pi}
{1\over \sqrt{\Lambda_c^2-m_0^2}}.
\eeq
If $\Lambda_c$ is taken to be infinity, the second term can be dropped.
For $s>m_0^2$, it is dominated by real $\chi$ production.
The final state depends on whether $\chi$ decay inside or ouside
the detector. For $s<m_0^2$ or $t$-channel exchange, the effective
interaction is
\beq
-{|y_f|^2\over 4 M^3}{1\over \sqrt{m_0^2-p^2}} |\bar{f}_L f_R h|^2.
\eeq
The singularity at $p^2=m_0^2$ will be smoothed out by the decay
width of the $\chi$ KK states.

There are also interactions with one or both $h$
replaced by their vev. For $\delta>2$,
the effective operators (\ref{EffOp}) are independent of
the momentum. They can be implemented
in COMPHEP via the exchange of an auxiliary field,
without the need to consider $t$-channel
and $u$-channel contributions separately. They are automatically
accounted for by COMPHEP in the process of constructing all
Feynman diagrams for the particular process.

\newpage

\vfill
\end{document}